\documentclass[11pt,a4paper]{article}
\pdfoutput=1

\usepackage{jheppub}
\usepackage{latexsym}
\usepackage{multirow}
\usepackage{color}
\usepackage[usenames,dvipsnames,table]{xcolor}
\usepackage{graphicx}
\usepackage{epsfig}  
\usepackage{epsf}    
\usepackage{dcolumn}
\usepackage{bm}
\usepackage{dcolumn}
\usepackage{textcomp}
\usepackage{float}
\usepackage{subfig}
\usepackage{hypcap}
\usepackage[]{hyperref}
\usepackage{makecell}
\usepackage{color}
\usepackage{pifont}
\usepackage{appendix}
  
\hypersetup{
  bookmarks=true,         
  unicode=false,          
  pdftoolbar=true,        
 pdfmenubar=true,        
 pdffitwindow=true,     
 pdfstartview={FitH},    
 pdfsubject={Neutrino Oscillations Phenomenology},   
 pdfnewwindow=true,      
 pdfcreator={RevTeX},
 colorlinks=true,       
 linkcolor=red,          
 citecolor=blue,        
 filecolor=black,      
 urlcolor=blue,           
  }

\newcommand{\be}{\begin{equation}}
\newcommand{\ee}{\end{equation}}
\newcommand{\ba}{\begin{eqnarray}}
\newcommand{\ea}{\end{eqnarray}}

\newcommand{\ms}{\Delta m^2_{21}}
\newcommand{\ma}{\Delta m^2_{31}}
\newcommand{\mam}{\Delta m^2_{32}}
\newcommand{\meff}{\Delta m^2_{\textrm{eff}}}

\newcommand{\ts}{\sin^22\theta}

\newcommand{\stch}{\sin^22\theta_{13}}
\newcommand{\sa}{\sin^2\theta_{23}}
\newcommand{\sta}{\sin^22\theta_{23}}

\newcommand{\dcp}{\delta_{\textrm{CP}}}
\newcommand{\tmt}{\theta_{23}}
\newcommand{\tet}{\theta_{13}}
\newcommand{\tem}{\theta_{12}}

\def\nue{{\nu_e}}
\def\anue{{\bar\nu_e}}
\def\numu{{\nu_{\mu}}}
\def\anumu{{\bar\nu_{\mu}}}

\def\chisqical{\chi^2_{\rm ~ICAL}}
\def\chisqmh{\Delta\chi^2_{\rm ~ICAL-MH}}
\def\chisqpm{\Delta\chi^2_{\rm ~ICAL-PM}}
\def\chisqos{\Delta\chi^2_{\rm ~ICAL-OS}}

\newcommand{\capdef}{}
\newcommand{\mycaption}[2][\capdef]{\renewcommand{\capdef}{#2}
       \caption[#1]{{\footnotesize #2}}}
\makeatletter
\renewcommand{\fnum@table}{\textbf{\tablename~\thetable}}
\renewcommand{\fnum@figure}{\textbf{\figurename~\thefigure}}
\makeatother

\preprint{IP/BBSR/2014-3, TIFR/TH/14-07}

\title{Enhancing sensitivity to neutrino parameters at INO 
combining muon and hadron information}

\author[a]{Moon Moon Devi,} 
\author[a,b]{Tarak Thakore,}
\author[c]{Sanjib Kumar Agarwalla,}
\author[a]{Amol Dighe} 

\affiliation[a]{Tata Institute of Fundamental Research, Mumbai 400005, India}
\affiliation[b]{Louisiana State University, Baton Rouge, Louisiana 70803, USA}
\affiliation[c]{Institute of Physics, Sachivalaya Marg, Sainik School Post, Bhubaneswar 751005, India}

\emailAdd{moonmoon4u@tifr.res.in}
\emailAdd{thakore@phys.lsu.edu}
\emailAdd{sanjib@iopb.res.in}
\emailAdd{amol@tifr.res.in}

\abstract
{
The proposed ICAL experiment at INO aims to identify the neutrino mass 
hierarchy from observations of atmospheric neutrinos, and help improve 
the precision on the atmospheric neutrino mixing parameters.
While the design of ICAL is primarily optimized to measure muon momentum,
it is also capable of measuring the hadron energy in each event.
Although the hadron energy is measured with relatively lower resolution,
it nevertheless contains crucial information on the event, which may be 
extracted when taken concomitant with the muon data.
We demonstrate that by adding the hadron energy information to the muon
energy and muon direction in each event, the sensitivity 
of ICAL to the neutrino parameters can be improved significantly.
Using the realistic detector response for ICAL, we present its enhanced
reach for determining the neutrino mass hierarchy, the
atmospheric mass squared difference and the mixing angle $\theta_{23}$,
including its octant. In particular, we show that the analysis that
uses hadron energy information can distinguish the normal and inverted 
mass hierarchies with $\Delta \chi^2 \approx 9$ with 10 years exposure
at the 50 kt ICAL, which corresponds to about 40\% improvement over the 
muon-only analysis. 
}

\arxivnumber{1406.3689}

\begin{document}
\maketitle
\flushbottom

\section{Introduction and Motivation}
\label{introduction}

After the recent discovery of a nonzero mixing angle $\tet$
at reactor $\bar{\nu}_e$ disappearance experiments 
\cite{An:2013zwz,An:2012bu,Ahn:2012nd,Abe:2011fz,Abe:2012tg} and
accelerator $\nu_e/\bar{\nu}_e$ appearance experiments
\cite{Adamson:2013ue,Abe:2013hdq,Abe:2013xua},  the two major remaining unknown issues
in neutrino oscillations are (i) whether the neutrino mass hierarchy (MH) is
normal (NH) or inverted (IH), i.e. whether $\Delta m^2_{32} \equiv m_3^2 - m_2^2$ is positive or negative,
respectively, and (ii) the possible presence of CP violation
\cite{Pascoli:2013wca,Agarwalla:2013hma}. 
Here $m_3$ corresponds to the neutrino mass eigenstate
with the smallest electron component.
The moderately large value of $\tet$ enables us to probe the 
sub-leading three-flavor effects in current and future neutrino 
oscillation experiments in order to address these unknowns \cite{Agarwalla:2014fva,Minakata:2014tza}.
In particular, the mass hierarchy, which is a very potent discriminator 
among models of neutrino mass generation \cite{Albright:2006cw}
can be probed through the measurement of matter effects 
\cite{Wolfenstein:1977ue,Wolfenstein:1979ni,Barger:1980tf,Mikheev:1986gs,Mikheev:1986wj} 
on neutrinos as they pass through the Earth over long distances. 
The matter effects induce characteristic differences in the neutrino and antineutrino signals
\cite{Blennow:2013rca,Smirnov:2013cqa}, which is the key to unravel the neutrino MH.

The race for the neutrino MH has received a tremendous boost after the discovery of a moderately 
large value of $\tet$. Looking at the current and future neutrino roadmap, a resolution of 
this issue certainly seems possible in coming ten years or so \cite{Blennow:2013oma}. 
Several experimental strategies have been adopted or proposed to determine the type of the neutrino MH.
Current generation off-axis long-baseline accelerator experiments T2K \cite{Itow:2001ee,Abe:2011ks} and 
NO$\nu$A \cite{Ayres:2002ws,Ayres:2004js,Ayres:2007tu} are expected to provide the first hint 
of neutrino MH \cite{Huber:2009cw,Agarwalla:2012bv} by observing the appearance of $\nue$ ($\anue$)
events in a $\numu$ ($\anumu$) beam. Future on-axis superbeam facilities consisting of intense, high power 
wide-band beams and large next generation detectors, like LBNE \cite{Diwan:2003bp,Barger:2007yw,Huber:2010dx,Akiri:2011dv,Adams:2013qkq}  
and LBNO \cite{Autiero:2007zj,Rubbia:2010fm,Angus:2010sz,Rubbia:2010zz,Agarwalla:2011hh,Stahl:2012exa,Agarwalla:2013kaa}
due to their relatively longer baselines, can settle this issue with much higher confidence. 
Another interesting test bed for the neutrino MH is the class of 
medium-baseline reactor experiments, like the proposed 
JUNO \cite{Li:2014qca} and RENO-50 \cite{RENO-50}.
These future facilities will discriminate between the two different 
MHs not by using the Earth's matter effect, but through the observation of the interference pattern between the
two oscillation frequencies in the reactor antineutrino energy spectrum.

Atmospheric neutrinos can also play a
crucial role in this direction. The precise study of atmospheric neutrinos at GeV energies traveling large distances 
is enriched with Earth's matter effects which in turn gives information on the neutrino 
MH \cite{Akhmedov:1998ui,Akhmedov:1998xq,Chizhov:1999az,Banuls:2001zn,Gandhi:2004md,Barger:2012fx}. 
The smallness of the atmospheric neutrino flux at GeV energies can be compensated by using very large detectors, 
like the low energy extension of IceCube, called PINGU \cite{Aartsen:2014oha} and within the context of the KM3NeT
project, a first phase with a dense detector in the open ocean, known as ORCA \cite{Katz:2014tta}. 
Recently, a lot of attention has been given to estimate the MH discovery potential of these proposed 
facilities \cite{Akhmedov:2012ah,Agarwalla:2012uj,Franco:2013in,Ribordy:2013xea,Winter:2013ema,Blennow:2013vta,Ge:2013zua,Ge:2013ffa}
in light of the large $\tet$. The proposed magnetized Iron Calorimeter (ICAL) detector located at the 
India-based Neutrino Observatory (INO) cavern \cite{INO,Athar:2006yb} is being designed to observe the 
atmospheric neutrinos at GeV energies with high detection efficiency and excellent energy and 
angular resolution for muons \cite{Chatterjee:2014vta,Agarwalla:NNN2013}. 
The most important feature of the ICAL detector is its charge identification capability using a 
magnetic field which makes it possible to observe $\numu$ and $\anumu$ events separately.
It gives the ICAL detector an edge compared to the other running or proposed atmospheric neutrino experiments
and greatly enhances the MH discovery reach without diluting the 
Earth's matter effect contained in neutrino and antineutrino 
signals \cite{PalomaresRuiz:2004tk,Indumathi:2004kd,Petcov:2005rv,Samanta:2006sj,Gandhi:2007td,Blennow:2012gj,Ghosh:2012px}.  

Though the main focus of ICAL is identification of the neutrino MH, 
it will also contribute to the precision measurements
of the atmospheric neutrino mixing parameters, 
viz. $|\Delta m^2_{32}|$ and $\theta_{23}$ \cite{Thakore:2013xqa}.
One of the major questions here, from the point of view of building
models of neutrino mass and mixing 
\cite{Mohapatra:2006gs,Albright:2006cw,Albright:2010ap,King:2013eh}
that try to explain the two large and one relatively small mixing angle 
in the lepton sector, is whether $\theta_{23}$ is maximal or not,
and if it is indeed non-maximal, whether $\theta_{23}$ is less than 
$45^\circ$ (the lower octant -- LO -- solution) or greater than 
$45^\circ$ (the higher octant -- HO -- solution). This is the so-called 
problem of octant degeneracy of $\tmt$ \cite{Fogli:1996pv},
which could also be addressed partly by ICAL.
This experiment would also be able to put severe constraints on
new physics scenarios like CPT violation \cite{Chatterjee:2014oda},
and will significantly enhance the reach of T2K and NO$\nu$A for 
detecting CP violation \cite{Ghosh:2013yon}.
 
ICAL is best suited for observing interactions of $\nu_\mu$ and $\bar{\nu}_\mu$ 
from the atmospheric neutrinos, which have energies in the GeV range.
When these neutrinos undergo charged-current interactions in the
detector, they give rise to muons, which are tracked by the
resistive plate chambers (RPCs) that constitute the active component
of the detector.  
The ICAL has been designed to efficiently detect muons
of energies in the GeV range, identify their charge, and reconstruct 
their momenta to a high precision \cite{Chatterjee:2014vta,Agarwalla:NNN2013}. 
The typical efficiency for detection of a 5 GeV muon travelling vertically is 80\%, 
while the typical charge identification efficiency is more than 95\%. 
The energy $E_\mu$ of such a muon can typically be reconstructed with an 
accuracy of 12\%, while its direction may be reconstructed to $1^\circ$ 
\cite{Chatterjee:2014vta,Agarwalla:NNN2013}. 
Owing to this capability, the initial analyses of the physics 
reach of ICAL have focused on the information from muon energy and
direction only \cite{Ghosh:2012px,Thakore:2013xqa,Chatterjee:2014oda}.

However one of the unique features of ICAL is its ability to detect
hadron showers and extract information about hadron energy and direction
from them. For example, the difference in energies of the interacting 
neutrino and the outgoing muon, $E'_{\rm had} \equiv E_\nu - E_\mu$, 
can be calibrated against the number of hits in the detector due to 
the hadron shower. The measured number of hits can then be used to 
reconstruct the fraction of energy of the incoming neutrino that is 
carried by the hadron. This may be achieved with an energy resolution of 
85\% (36\%) for the hadron energy of 1 GeV (15 GeV) \cite{Devi:2013wxa}.
Though the achievable precision on $E'_{\rm had}$ is much lower than that 
on $E_\mu$, it still provides additional information about the particular event, 
which can be extracted in order to improve the physics reach of the detector.
Note that, it is quite challenging to extract the hadronic information at multi-GeV energies 
in currently running or upcoming water or ice based atmospheric neutrino detectors.

In ICAL, one way of using the hadron information would be to simply add the 
reconstructed values of $E_\mu$ and $E'_{\rm had}$ to reconstruct 
the energy of the incoming neutrino in each event, which indeed can improve the 
accuracy in the measurement of $|\Delta m^2_{32}|$
\cite{Samanta:2010xm,DU-preparation}\footnote{Such a reconstruction of incoming neutrino 
energy in multi-GeV range becomes quite difficult in the detectors like Super-Kamiokande due 
to the poor reconstruction efficiency of multi-ring events; this can be done 
with a high efficiency only in the sub-GeV range where single-ring events dominate.}. 
However in the process of adding 
$E_\mu$ and $E'_{\rm had}$ in ICAL, the advantage of high precision in the 
measurement of $E_\mu$ is partially lost in case of MH discrimination. 
It has been claimed in \cite{Ghosh:2013mga} that the MH discovery reach can be 
improved by treating the reconstructed muon momentum and calibrated $E'_{\rm had}$
as two separate variables. However, since the fraction of neutrino energy carried 
by the muon, or equivalently the inelasticity $y \equiv E'_{\rm had}/E_\nu$, 
is different for each event, the correlation
between these quantities constitutes an important part of the information about the event 
that should not be missed. This strategy has been suggested earlier in the context of the PINGU 
and ORCA experiments in \cite{Ribordy:2013xea}, where it has been pointed out that by exploring 
the information on the inelasticity parameter in each event,
the MH reach can be improved by 20-50\%.
We implement the same idea here in detail in the context of the ICAL 
experiment to enhance its MH discrimination capability as well as 
the precision on the atmospheric parameters.

We therefore adopt the approach of using the values of $E_\mu, 
\cos\theta_\mu$, and $E'_{\rm had}$ from each event as independent and
correlated pieces of information. In this study, we bin 
the data in all these three quantities, as opposed to 
the analyses that use only the muon momentum $(E_\mu, \cos\theta_\mu)$.
Of course, this also means that the already sparse data has to be further divided into 
a larger number of bins. Hence we choose to use a slightly coarser binning
for $E_\mu$ and $\cos\theta_\mu$.
As will be seen from the results, our approach results in a
marked improvement in the ability of the detector to identify the mass 
hierarchy and increase in the precision on $|\Delta m^2_{32}|$.
The magnitude of the improvement is of the same order as was 
expected in \cite{Ribordy:2013xea}.

The structure of the paper is as follows. 
In Sec.~\ref{sec:methodology} we outline our methodology: extraction of the 
hadron energy information, the binning scheme, and the $\chi^2$ procedure.
Sec.~\ref{sec:results} presents the results for the neutino mass hierarchy,
precision measurements of the atmospheric oscillation parameters, and for the
$\theta_{23}$ octant sensitivity. 
We conclude in Sec.~\ref{sec:conclusions} with a summary of results
and comments on our analysis.

\section{Methodology}
\label{sec:methodology}

\subsection{Neutrino Interactions and Event Reconstruction}

The ICAL detector, as described in \cite{INO,Agarwalla:NNN2013},
consists of alternate layers of iron plates and RPCs, which act as 
the target mass and active detection elements, respectively.
When a charged particle passes through an RPC, the (X,Y) coordinate of 
its path is recorded in the form of strip hits. 
The Z-coordinate is provided by the RPC layer number. The hits created by 
muons in a charged-current $\nu_\mu$ interaction give rise to distinct 
track-like features, while the hits created by hadrons produce shower-like 
features. 

\begin{figure}[h]
\centering
\includegraphics[width=0.49\textwidth]{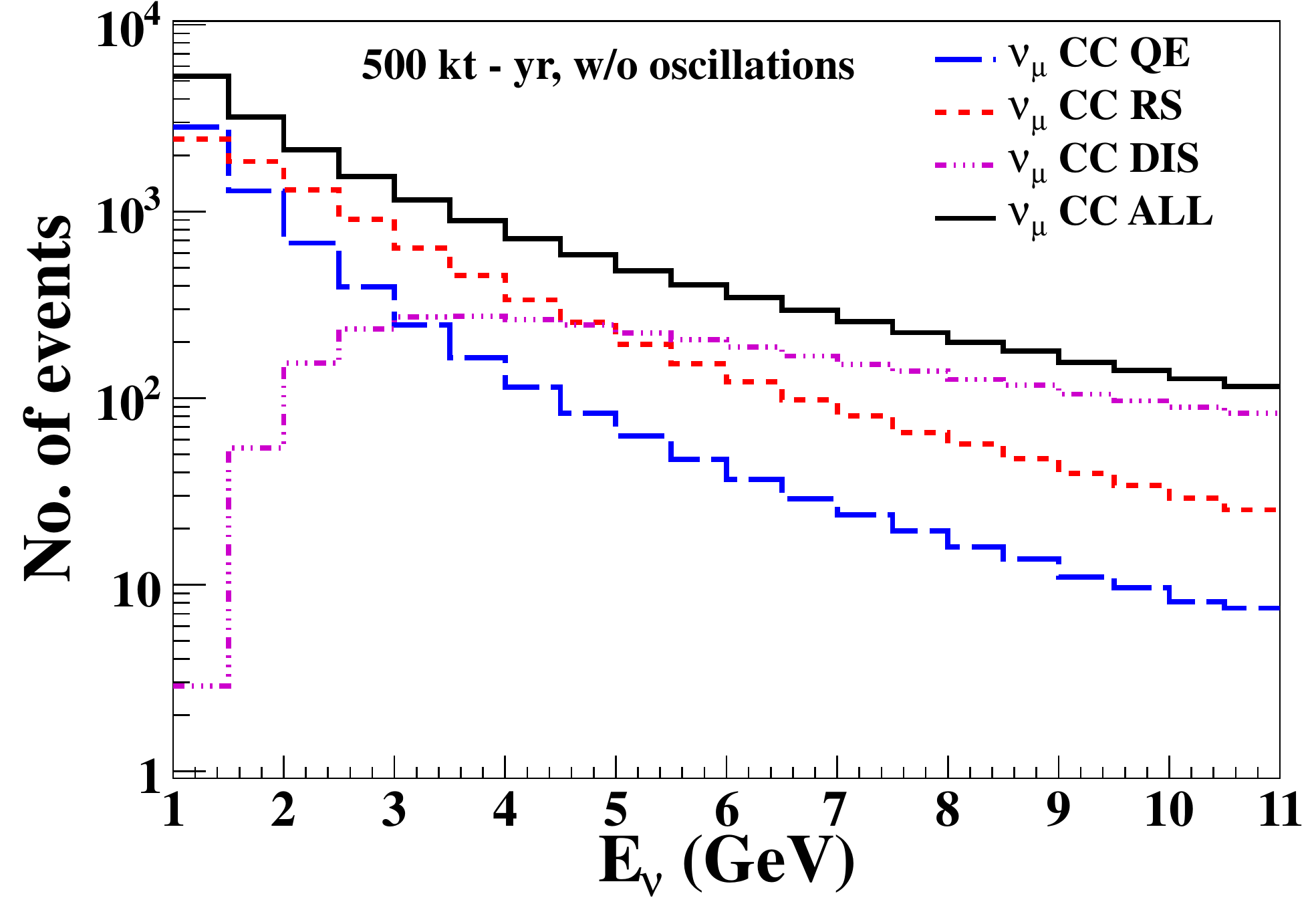}
\includegraphics[width=0.49\textwidth]{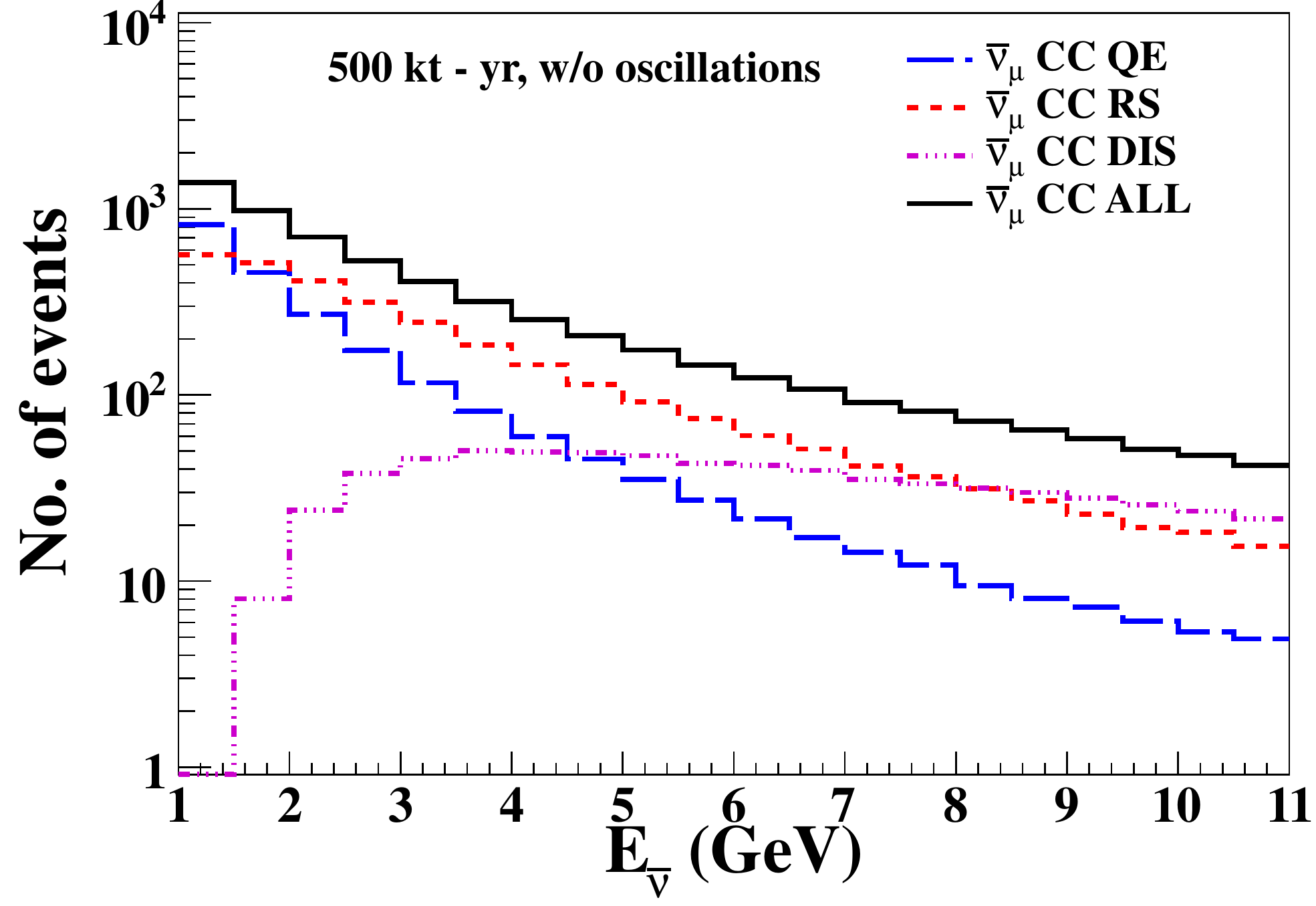}
\mycaption{The number of events in the QE, RS and DIS processes 
at ICAL, as functions of neutrino and antineutrino energies, with an exposure 
of 500 kt-yr, in the absence of oscillations. The total number of events
is also shown.}
\label{fig:nuance-events}
\end{figure}

\begin{figure}[t]
\centering
\includegraphics[width=0.49\textwidth]{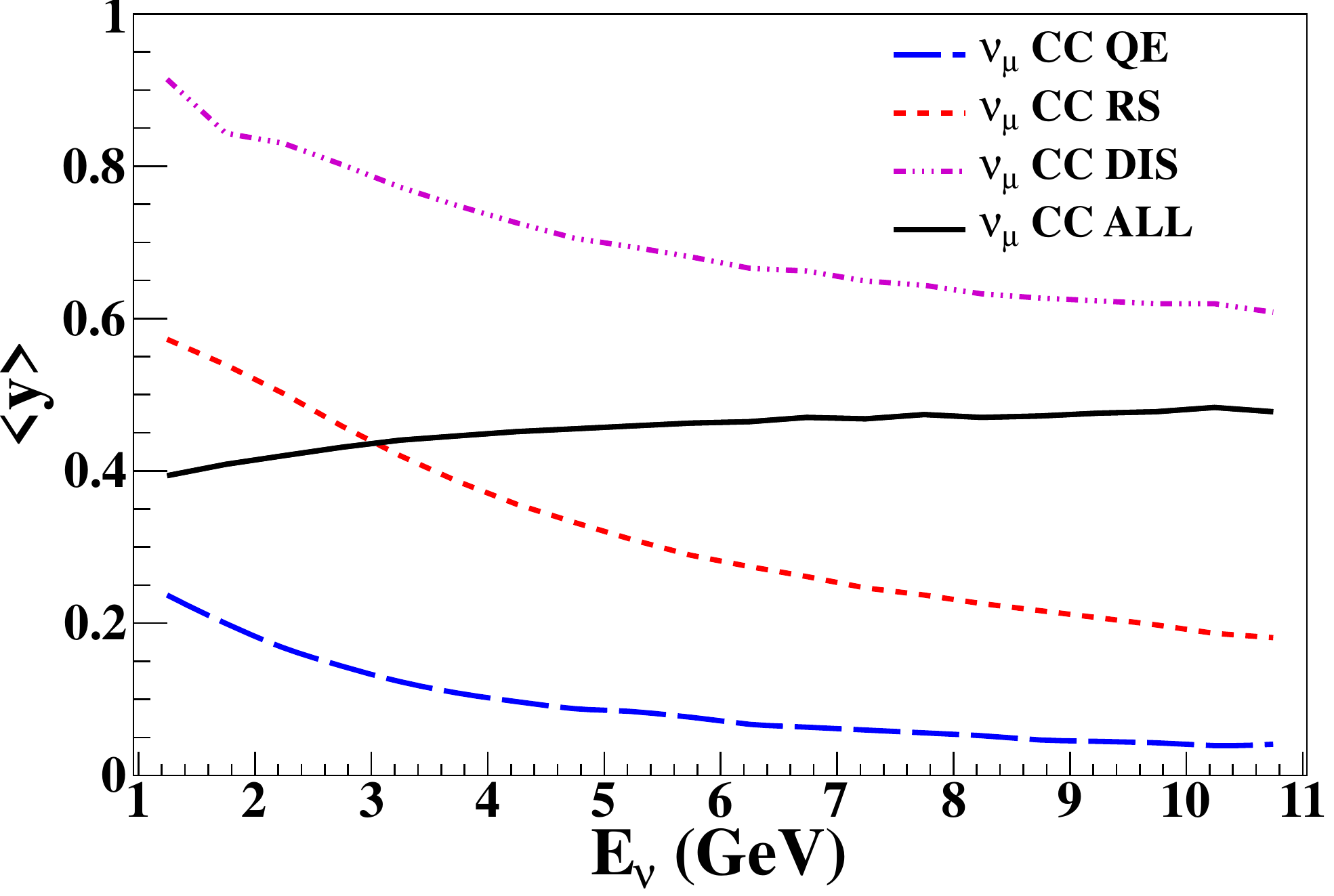}
\includegraphics[width=0.49\textwidth]{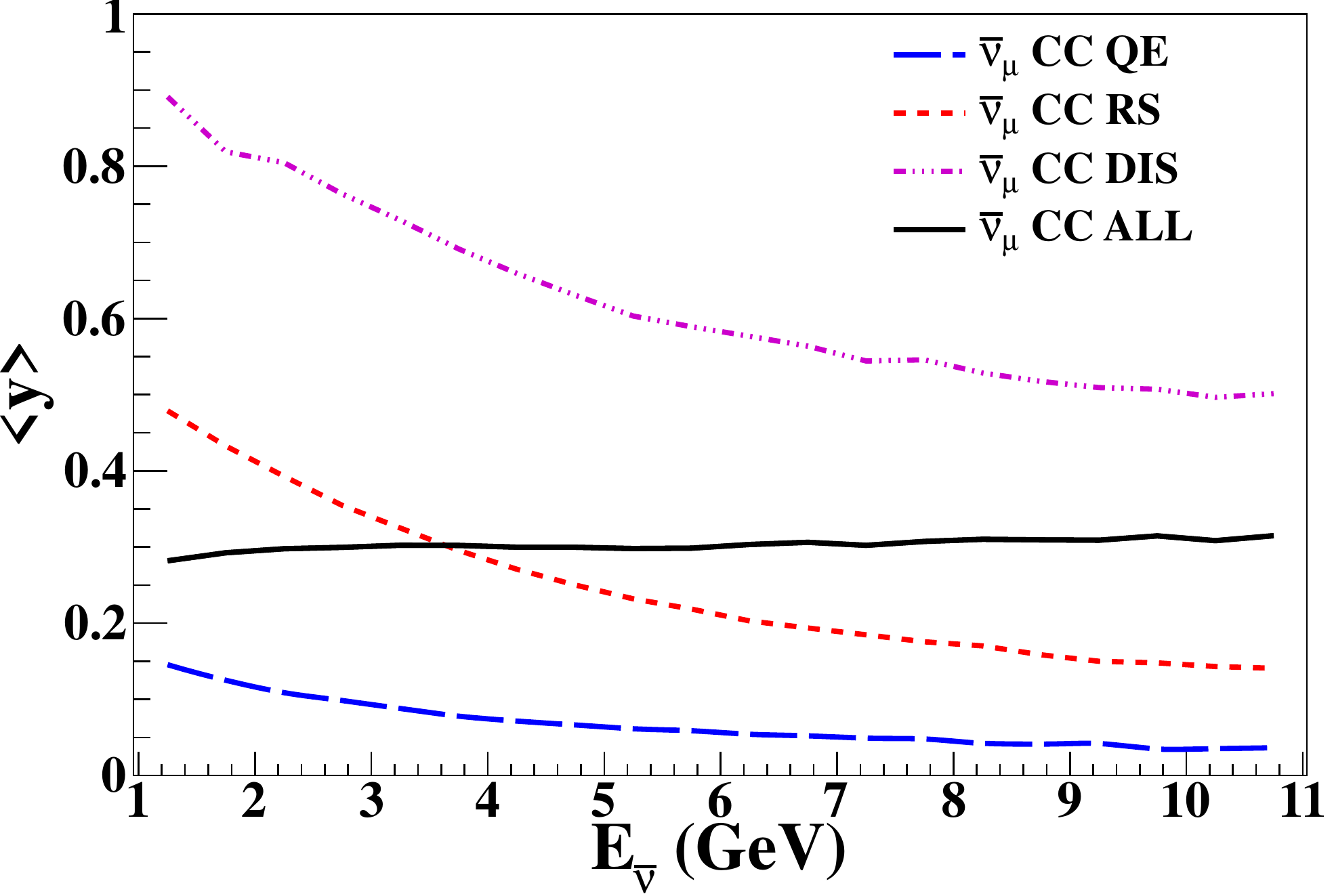}
\mycaption{The average inelasticities $\langle y \rangle$ 
in QE, RS, and DIS processes as a functions of neutrino and antineutrino 
energies. We also show $\langle y \rangle$ for all channels.}
\label{fig:y-avg}
\end{figure}

\begin{figure}[h]
\centering
\includegraphics[width=0.49\textwidth]{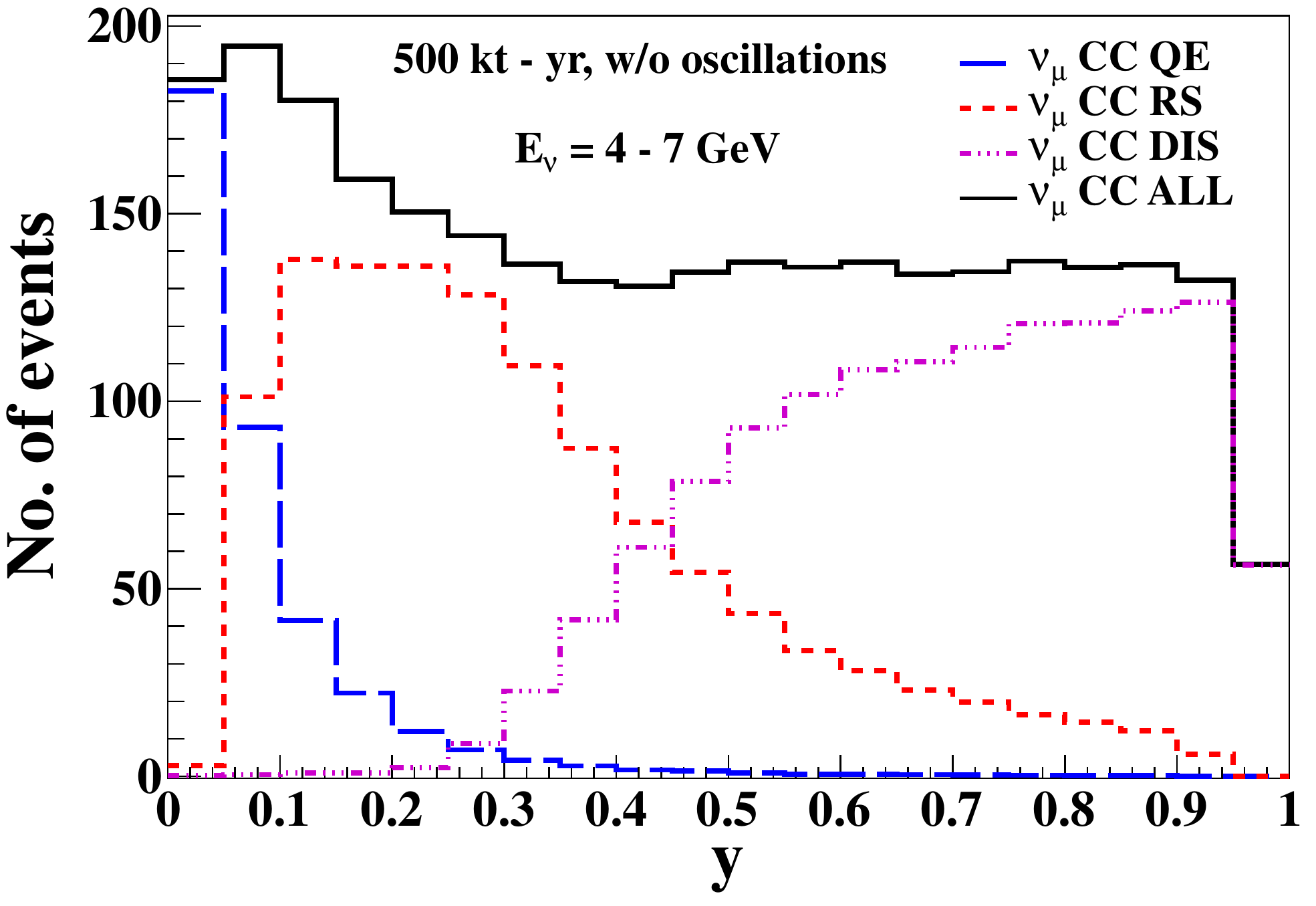}
\includegraphics[width=0.49\textwidth]{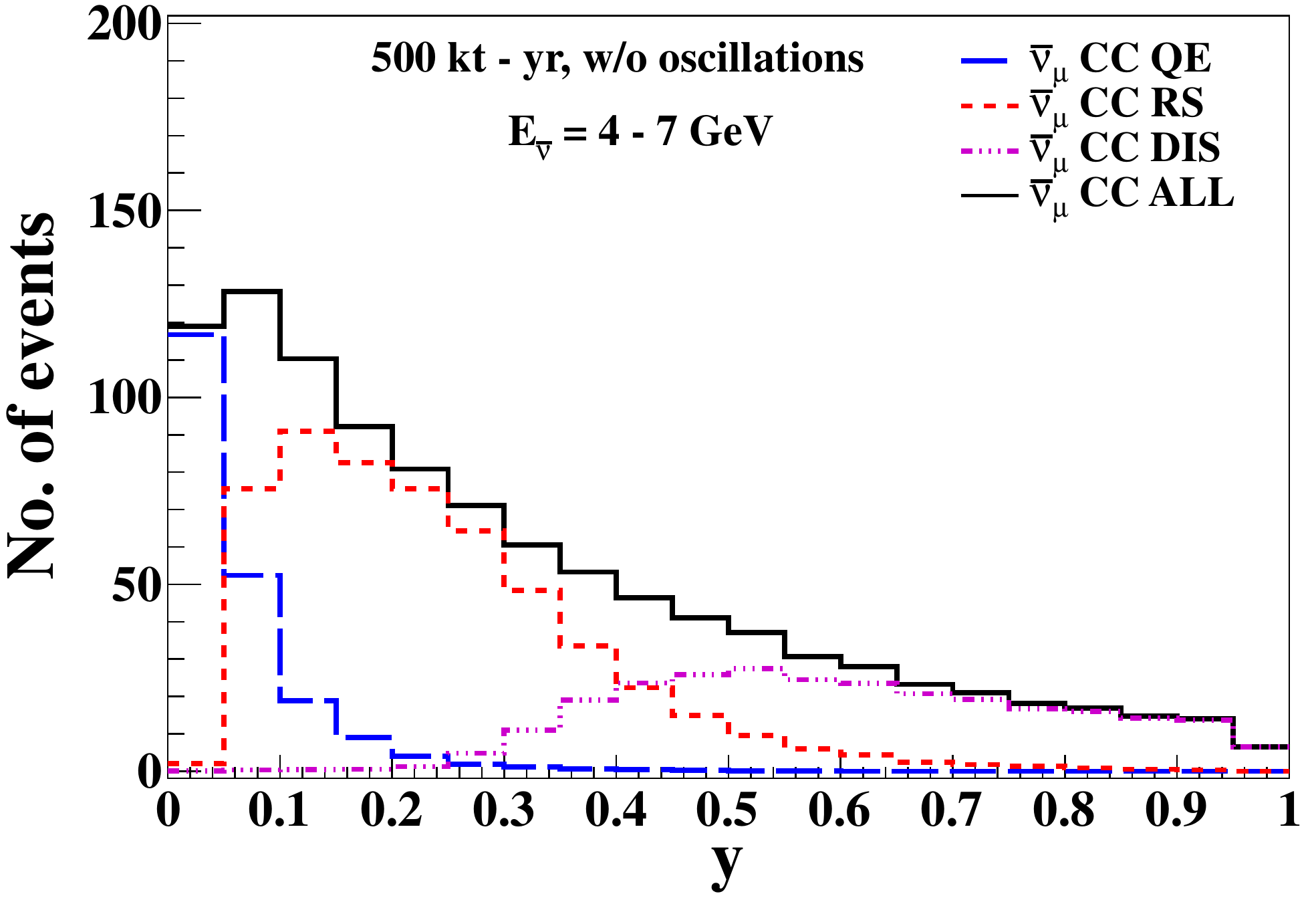}
\mycaption{The distribution of inelasticity in events with
neutrino and antineutrino energies in the range 4 to 7 GeV, with an
exposure of 500 kt-yr, in the absence of oscillations.}
\label{fig:y-distribution}
\end{figure}

Three main processes contribute to the charged-current $\nu_\mu$ interactions 
in the ICAL detector.
In the sub-GeV energy range of neutrinos, the quasi-elastic (QE) process 
dominates, where the final state muon carries most of the available energy 
and no hadrons are produced. Hadronic showers make their appearance
in resonance (RS) and deep-inelastic scattering (DIS) processes when we move 
from sub-GeV to multi-GeV range. In the RS process, the final state hadron 
shower mostly consists of a single pion, though multiple pions may contribute
in a small fraction of events. The DIS process produces multiple hadrons, 
which carry a large fraction of the incoming neutrino energy.
Fig.~\ref{fig:nuance-events} shows the relative contributions of these three 
processes to the total number of events in the absence of oscillations, 
obtained using the event generator NUANCE \cite{Casper:2002sd} 
and the atmospheric neutrino fluxes at Kamioka \cite{Honda:2011nf}
that we also use in our further analysis in this paper\footnote{
Note that the figures \ref{fig:nuance-events}, \ref{fig:y-avg}, and
\ref{fig:y-distribution} are drawn at the generator level and include the 
information on the cross section with the target. The detector 
efficiencies and resolutions are not included.}.
It may be observed that in the neutrino energy range of 5 to 10 GeV,
the contribution of DIS events is significant. This is precisely the energy range
where one expects significant matter effects that will help the mass hierarchy 
identification.  The information on hadrons produced in these DIS events is 
therefore crucial.

The inelasticity in an event, defined as 
$y \equiv (E_\nu-E_\mu)/E_\nu = E'_{\rm had}/E_\nu$, 
is roughly the fraction of the neutrino energy that is carried by hadrons. 
The average inelasticities  $\langle y \rangle$ in the three kinds of
processes, as functions of neutrino and antineutrino energies, 
have been shown in Fig.~\ref{fig:y-avg}. Clearly, the average inelasticity
in the DIS events is significant, which implies that in the energy range of
interest for mass hierarchy identification, a large fraction of
the incoming neutrino energy goes into hadrons. While the average inelasticity
does not vary much over this energy range, the inelasticities in individual
events have a wide distribution (see Fig.~\ref{fig:y-distribution}). 
Hence it is important to take into account the $y$ values in individual events. 
In this paper, we use the energies of hadrons and muons obtained in each event 
individually, so that the correlation between them is preserved.

\begin{figure}[h]
\centering
\includegraphics[width=0.75\textwidth]{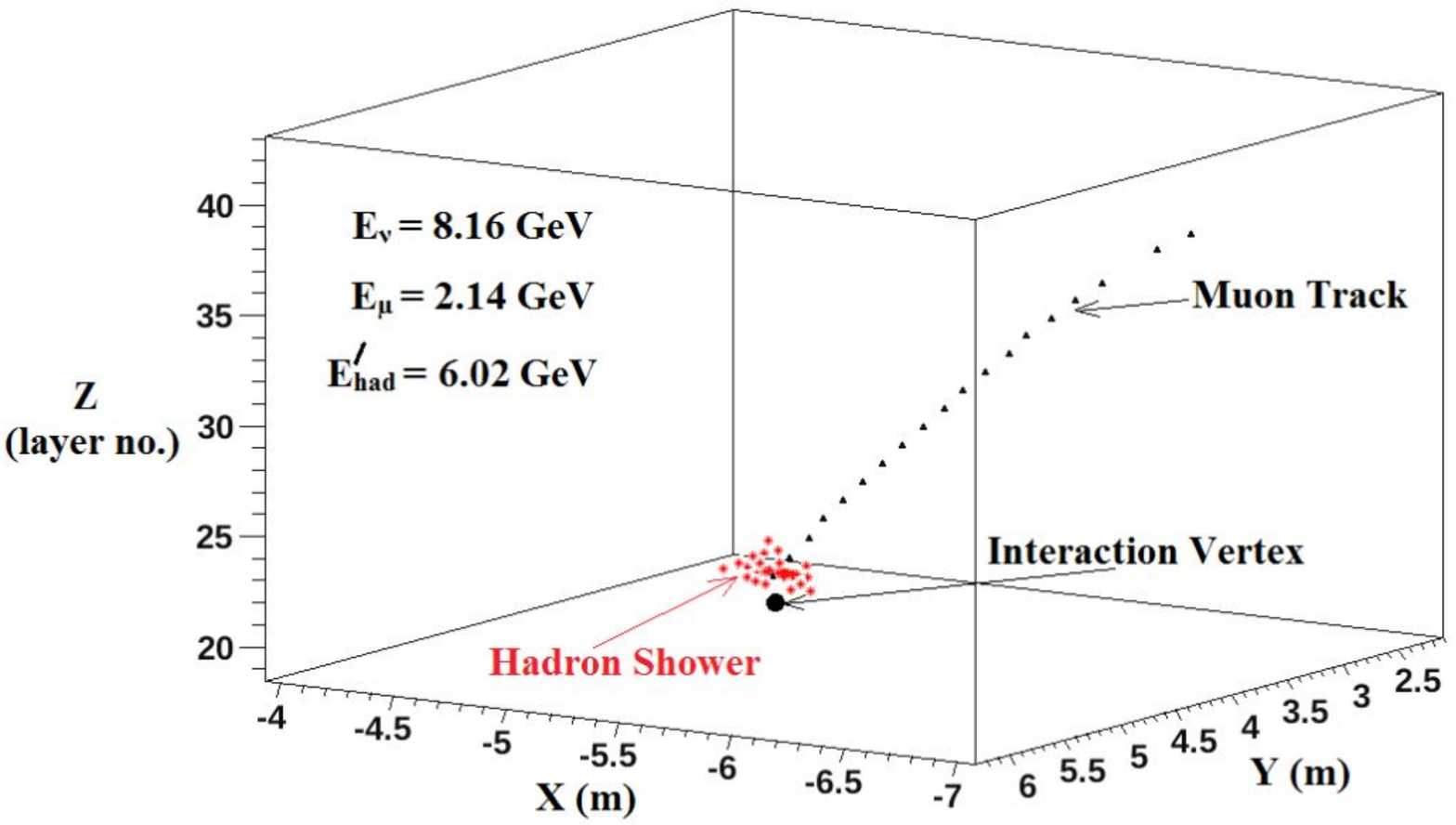}
\mycaption{A typical deep-inelastic atmospheric 
muon-neutrino event in the ICAL detector,
obtained using the GEANT4 simulation. Only the 
relevant part of the detector is shown. X and Y denote length 
in units of meters whereas Z represents the layer number.}
\label{event-display}
\end{figure}

We have already mentioned that distinct tracks are created by muon hits and 
shower-like features emerge from the hadron hits.
Figure~\ref{event-display} illustrates a neutrino 
interaction in the simulated ICAL detector (drawn using the VICE event 
display package \cite{vice}) producing a muon track and a hadron shower.
The muon reconstruction for ICAL is described in 
\cite{Chatterjee:2014vta}, while the hadron reconstruction is
described in \cite{Devi:2013wxa}.

In this work, we focus only on the charged-current event where the
neutrino interaction produces a muon and possibly also a hadron shower.
Note that in general, it may not be always possible to distinguish between
the muon track and the hadron shower in all events. Here, we assume that the 
hits created by a muon and hadron can be 
separated with 100\% efficiency by the ICAL particle reconstruction 
algorithms. This indeed was also the assumption made while obtaining
the muon and hadron response in \cite{Chatterjee:2014vta,Devi:2013wxa}. 
Whenever a muon is reconstructed, we take all the other hits to be 
a part of the hadronic shower for the purpose of hadron energy calibration. 
This is consistent with the procedure used in \cite{Devi:2013wxa} for 
determining the hadron energy calibration. This further implies that 
the neutrino event reconstruction efficiency is the same as the muon
reconstruction efficiency. Note that, the calibration of $E'_{\rm had}$ 
against the number of hadron shower hits also allows for the possibility of 
no hits observed in the hadron shower.
Finally, the background hits coming from other sources such as the
neutral-current events, charged-current $\nu_e$ events, cosmic muons, 
and the noise, have not been taken into account so far\footnote{ 
At a magnetized iron neutrino detector (MIND) which is similar to ICAL, 
the background due to neutral-current events and charged-current $\nu_e$ 
events can be reduced to the level of about a per cent by using the cuts 
on track quality and kinematics \cite{Bayes:2012ex}.}.
The systematics due to these effects will have to be taken care of
in future, as the understanding of the ICAL detector improves.

\subsection{Binning scheme in ($E_\mu$--$\cos\theta_\mu$--$E'_{\rm had}$) space}
\label{sec:binning}

After incorporating the reconstruction efficiencies and 
resolutions for muons and hadrons, in the absence of oscillations 
one would get about 6200 events with a $\mu^-$ and 2800 events
with a $\mu^+$, for an exposure of 500 kt-yr. 
These numbers would decrease further with oscillations.
For the analysis presented in~\cite{Ghosh:2012px}, 20 uniform $E_\mu$ bins 
in the range 1 to 11 GeV and 80 uniform $\cos\theta_\mu$ bins in the 
range $[-1,+1]$ were used for each polarity of muon. 
While the excellent energy and angular resolutions of muon in ICAL
\cite{Chatterjee:2014vta} allow us to use such a fine binning scheme, 
it does not ensure sufficient statistics for many bins.
Including $E'_{\rm had}$ as an additional observable for binning 
would increase the total number of bins further, reducing the statistical 
strength of each bin significantly.
To avoid this situation we use a coarser binning scheme that is suitable 
for the three observables $E_\mu$, $\cos\theta_\mu$, and $E'_{\rm had}$.
Now most of the bins have sufficient statistics while at the same time
the results are not diluted substantially. 

The optimized binning scheme would depend on the parameters one wants to
measure. In particular, it could be different for the mass hierarchy
identification and precision measurements of atmospheric neutrino
mixing parameters. We do not perform an optimization study for binning in this
paper, however we identify the regions in the 3-dimensional parameter
space ($E_\mu$--$\cos\theta_\mu$--$E'_{\rm had}$) that are sensitive 
to the mass hierarchy, and use finer binning in those regions.
These regions roughly span the intervals for $E_\mu$ = 4 to 7 GeV, 
$\cos\theta_\mu$ = -1 to -0.4, and $E'_{\rm had}$ = 0 to 4 GeV.
We use coarser binning in other regions.
The atmospheric neutrino flux follows a steep power law,
resulting in a smaller number of events at higher muon and hadron energies. 
Therefore, in general, we take finer bins at low energies and wider bins at 
higher energies, for both muons and hadrons, to ensure sufficient statistics 
in each bin. This is also consonant with larger uncertainties in
energy measurement at higher energies.
Our binning scheme is given in Table~\ref{table:3d-bin}.
For each polarity, we use 10 bins for $E_\mu$, 21 bins for $\cos\theta_\mu$, 
and 4 bins for $E'_{\rm had}$, resulting into a total of 
($4 \times 10 \times 21$) = 840 bins per polarity.

\begin{table}[t] 
\centering 
\begin{tabular}{|c| c| c| c|} 
\hline\hline 
Observable & Range & Bin width & Total bins \\ 
\hline
$E_{\mu}$ (GeV) & \makecell[c]{ $[1,4]$ \\ $[4,7]$ \\ $[7,11]$} & 
\makecell[c]{0.5 \\ 1 \\4} & 
$\left.\begin{tabular}{l}
6 \\ 3 \\ 1
\end{tabular}\right\}$  10\\
\hline
$\cos\theta_\mu$  & \makecell[c]{ $[-1.0,-0.4]$ \\ $[-0.4,0.0]$ \\ $[0.0,1.0]$} 
& \makecell[c]{0.05 \\ 0.1 \\0.2} & 
$\left.\begin{tabular}{l}
12 \\ 4 \\ 5
\end{tabular}\right\}$  21\\
\hline
$E'_{\rm had}$ (GeV)  & \makecell[c]{$[0,2]$ \\ $[2,4]$ \\ $[4,15]$} 
& \makecell[c]{1 \\ 2 \\11} & 
$\left.\begin{tabular}{l}
2 \\ 1 \\ 1
\end{tabular}\right\}$  4\\
\hline\hline
\end{tabular}
\mycaption{The binning scheme adopted for the reconstructed 
observables $E_\mu$, $\cos\theta_\mu$, and $E'_{\rm had}$ 
for each muon polarity. The last column shows the total number 
of bins taken for each observable.}
\label{table:3d-bin}
\end{table}

\subsection{Details of the numerical analysis}
\label{sec:numerical}

In this analysis, we obtain the physics reach of the ICAL experiment by 
suppressing the statistical fluctuations of the ``observed'' event 
distribution which are simulated by NUANCE in the 
event rates as well as in the event kinematics. This is implemented 
by generating events for an exposure of 50,000 kt-yr, followed by 
incorporating the detector response and then normalizing the event 
distribution to the actual exposure. This procedure, with the
$\chi^2$ statistics, is expected to give the median sensitivity of 
the experiment in the frequentist approach \cite{Blennow:2013oma}. 

For event generation and inclusion of oscillation, we use the same 
procedure as described in \cite{Ghosh:2012px,Thakore:2013xqa},
provisionally using the neutrino fluxes predicted at Kamioka 
\cite{Honda:2011nf}\footnote{First calculations of the expected 
fluxes at the INO site have recently become available \cite{Athar:2012it}, 
and will be implemented in future analysis once they are finalized.
The difference in the fluxes at the INO and Kamioka sites arises from
the different horizontal components of geomagnetic field at these sites
(40$\mu$T  at the INO, 30$\mu$T at Kamioka).}. 
The detector response for muons is incorporated by smearing the
true muon energy and direction using the Gaussian distributions
with the resolution functions obtained from the ICAL detector
simulation \cite{Chatterjee:2014vta}.
The efficiencies of reconstruction and charge identification of muons
are also incorporated using the procedure described therein.
The hadron energy response is similarly incorporated by smearing 
the true hadron energy using the Vavilov distribution with the
parameters obtained from the fits to the ICAL detector simulations
\cite{Devi:2013wxa}.
After incorporating the detector response for muons and 
hadrons, for the true values of the oscillation parameters 
as given in Table~\ref{table:benchpar} (with $\stch$ = 0.1,
$\sa$ = 0.5, and NH), one gets about 4500 events with $\mu^-$ and 
about 2000 events with $\mu^+$ for a 500 kt-yr exposure. 
We obtain the distribution of these events in terms of 
$E_\mu$, $\cos\theta_\mu$, and $E'_{\rm had}$. 

We define the Poissonian $\chi^2_{-}$ for $\mu^{-}$ events as :
\begin{equation}
\chi^{2}_{-}
        ={\min_{\xi_l}} \sum_{i=1}^{N_{E'_{\rm had}}} \sum_{j=1}^{N_{E_{\mu}}} 
\sum_{k=1}^{N_{\cos \theta_{\mu}}} \left[ 2(N_{ijk}^{\rm theory} -  N_{ijk}^{\rm data})
        - 2 N_{ijk}^{\rm data} \: \ln \left( \frac{N_{ijk}^{\rm theory}}
{N_{ijk}^{\rm data}} \right) \right]
          + \sum_{l=1}^{5} \xi_{l}^{2}\,,
\label{chisq}
\end{equation}
where
\begin{equation}
N^{\rm theory}_{ijk} = N^{0}_{ijk}\bigg(1 + \sum_{l=1}^{5} \pi_{ijk}^{l} \xi_{l}\bigg)\,.
\label{n-theory-definition}
\end{equation}
In Eq.~(\ref{chisq}), $N_{ijk}^{\rm theory}$ and $N_{ijk}^{\rm data}$ are the
expected and observed number of $\mu^-$ events in a given ($E_{\mu}$,
$\cos \theta_{\mu}$, $E'_{\rm had}$) bin. 
$N^{0}_{ijk}$ are the number of events without systematic errors.
Here $N_{E_{\mu}}$ = 10, $N_{\cos\theta_{\mu}}$ = 21, and $N_{E'_{\rm had}}$ = 4,
as mentioned in Table~\ref{table:3d-bin}. To simulate $N_{ijk}^{\rm data}$,
we have used the oscillation parameters given in Table~\ref{table:benchpar} as
``true'' values. These are benchmark values used in our analysis, 
and are consistent with those allowed
by the global fit \cite{Forero:2014bxa,NuFIT,GonzalezGarcia:2012sz,Capozzi:2013csa}.
The effective mass-squared difference is related to the $\ma$ and $\ms$ mass-squared differences
through the expression~\cite{deGouvea:2005hk, Nunokawa:2005nx}:
\be
\meff = \ma - \ms (\cos^2\tem - \cos\dcp \, \sin\tet \, \sin2\tem \,
\tan\tmt) \, .
\label{parkedef}
\ee
The following five systematic errors are included in the analysis
using the method of pulls as in \cite{Ghosh:2012px,Thakore:2013xqa}: 
(i) Flux normalization error (20\%), (ii) cross-section error (10\%), 
(iii) tilt error (5\%), (iv) zenith angle error (5\%), and 
(v) overall systematics (5\%).

\begin{table}[t]
\begin{center}
\begin{tabular}{|c|c|c|} \hline
Parameter & True value & Marginalization range \\
\hline 
$\stch$ &  0.09, 0.1, 0.11 & [0.07, \,0.11]\cr
\hline
$\sa$ &  0.4, 0.5, 0.6 & [0.36, \,0.66] \cr
\hline
$\meff$ & $ \pm 2.4 \times 10^{-3} \ {\rm eV}^2$ & $[2.1, \, 2.6] \times 10^{-3} \ {\rm eV}^2$ (NH) \cr
& & $- [2.6, \, 2.1] \times 10^{-3} \ {\rm eV}^2$ (IH)  \cr
\hline
$\ts_{12}$ & 0.84 & Not marginalized  \cr
\hline$\ms$ & $ 7.5 \times 10^{-5} \ {\rm eV}^2$ & Not marginalized \cr
\hline
$\dcp$ & $ 0^{\circ}$ & Not marginalized \cr
\hline 
\end{tabular}
\mycaption{Benchmark oscillation parameters used in this analysis.
The second column shows the true values of the oscillation parameters used to 
simulate the ``observed" data set, where the ``true value'' is the choice 
of the parameter value for which the data is simulated.
The third column shows the range over
which the parameter values are varied while minimizing the $\chi^2$.
This range corresponds to the $3\sigma$ allowed values of the parameter 
in the global fit \cite{Forero:2014bxa,NuFIT,GonzalezGarcia:2012sz,Capozzi:2013csa}.
While performing the analysis for precision measurements in 
Sec.~\ref{PM-results}, we do not marginalize over $\meff$ or 
$\sin^2 \theta_{23}$, and take $|\mam{\rm (true)}|= 2.4 \times 10^{-3}$ eV$^2$.}
\label{table:benchpar}
\end{center}
\end{table}

Following an identical procedure, $\chi^2_{+}$ for $\mu^+$ events is obtained.
Total $\chi^2$ is obtained by adding the individual contributions from $\mu^-$ and 
$\mu^+$ events. We also add a 8\% prior (at 1$\sigma$) on $\sin^2 2\theta_{13}$, since
this quantity is currently known to this accuracy \cite{An:2013zwz,An:2012bu}. 
We do not use any prior on $\theta_{23}$ or $\Delta m^2_{32}$ since these 
parameters\footnote{Adding constraints on these parameters from other experiments 
will increase the global sensitivity to MH.} will be directly measured at the ICAL detector. 
Thus we define:

\begin{equation}
\chi^2_{\rm ~ICAL} =  \chi^{2}_{-} +  \chi^{2}_{+} + \chi^2_{\rm prior} \,,
\label{total-chisq}
\end{equation}
\begin{equation}
\chi^2_{\rm prior} \equiv \left(\frac{
\sin^2 2\theta_{13}- \sin^2 2\theta_{13}{\rm (true)}}
{\sigma(\stch)} \right)^2 \; . 
\end{equation}
We take $\sigma(\stch)$ = 0.08 $\times$ $\sin^2 2\theta_{13}{\rm (true)}$.
While implementing the minimization procedure, $\chisqical$ is first minimized 
with respect to the pull variables $\xi_l$, 
and then marginalized over the ranges of oscillation parameters 
$\sin^2 \theta_{23}$, $\Delta m^2_{\rm eff}$ and $\sin^2 2\theta_{13}$
as given in Table~\ref{table:benchpar}, wherever appropriate. 
We do not marginalize over $\delta_{\rm CP}, \Delta m^2_{21}$ and
$\theta_{12}$ since they have negligible effect on the relevant
oscillation probabilities at ICAL \cite{Akhmedov:2004ny}.
While we use the best-fit values of $\Delta m^2_{21}$ and
$\theta_{12}$ from the global fit references \cite{Forero:2014bxa,NuFIT,GonzalezGarcia:2012sz,Capozzi:2013csa},
we consider $\delta_{\rm CP}=0$ throughout our analysis.

\section{Results with the $(E_\mu, \cos\theta_\mu, E'_{\rm had})$ analysis}
\label{sec:results}

In this section, we discuss our findings. We first begin with addressing the neutrino
mass hierarchy issue. 

\subsection{Identifying the neutrino mass hierarchy}
\label{MH-results}

We quantify the statistical significance of the analysis to rule out the
wrong hierarchy by
\begin{equation}
\Delta \chi^2_{\rm ~ICAL-MH} = \chi^2_{\rm ~ICAL} (\text{false MH}) - 
\chi^2_{\rm ~ICAL} (\text{true MH}) .
\label{mh_chi2_def}
\end{equation}
Here, $\chisqical(\text{true MH})$ and $\chisqical(\text{false MH})$
are obtained by performing a fit to the ``observed'' data assuming
true and false mass hierarchy, respectively.
Here with the statistical fluctuations suppressed, 
$\chi^2_{\rm ~ICAL} (\text{true MH}) \approx 0$.
The statistical significance is also represented in terms of $n\sigma$,
where $n \equiv \sqrt{\chisqmh}$.
It has been demonstrated recently \cite{Blennow:2013oma}
that this relation gives the median sensitivity in the frequentist approach 
of hypothesis testing.

\begin{figure}[t]
\centering
\includegraphics[width=0.49\textwidth]{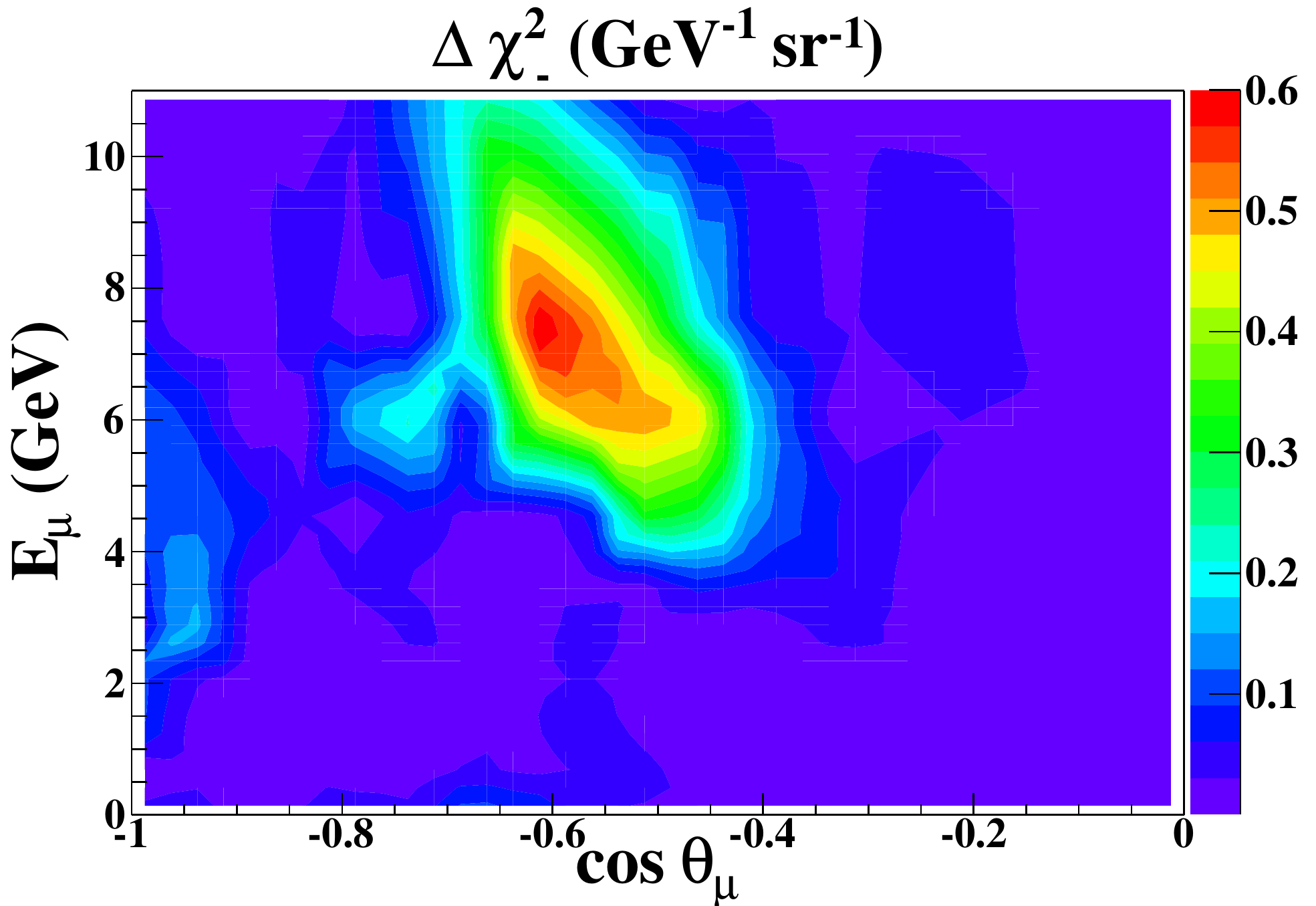}
\includegraphics[width=0.49\textwidth]{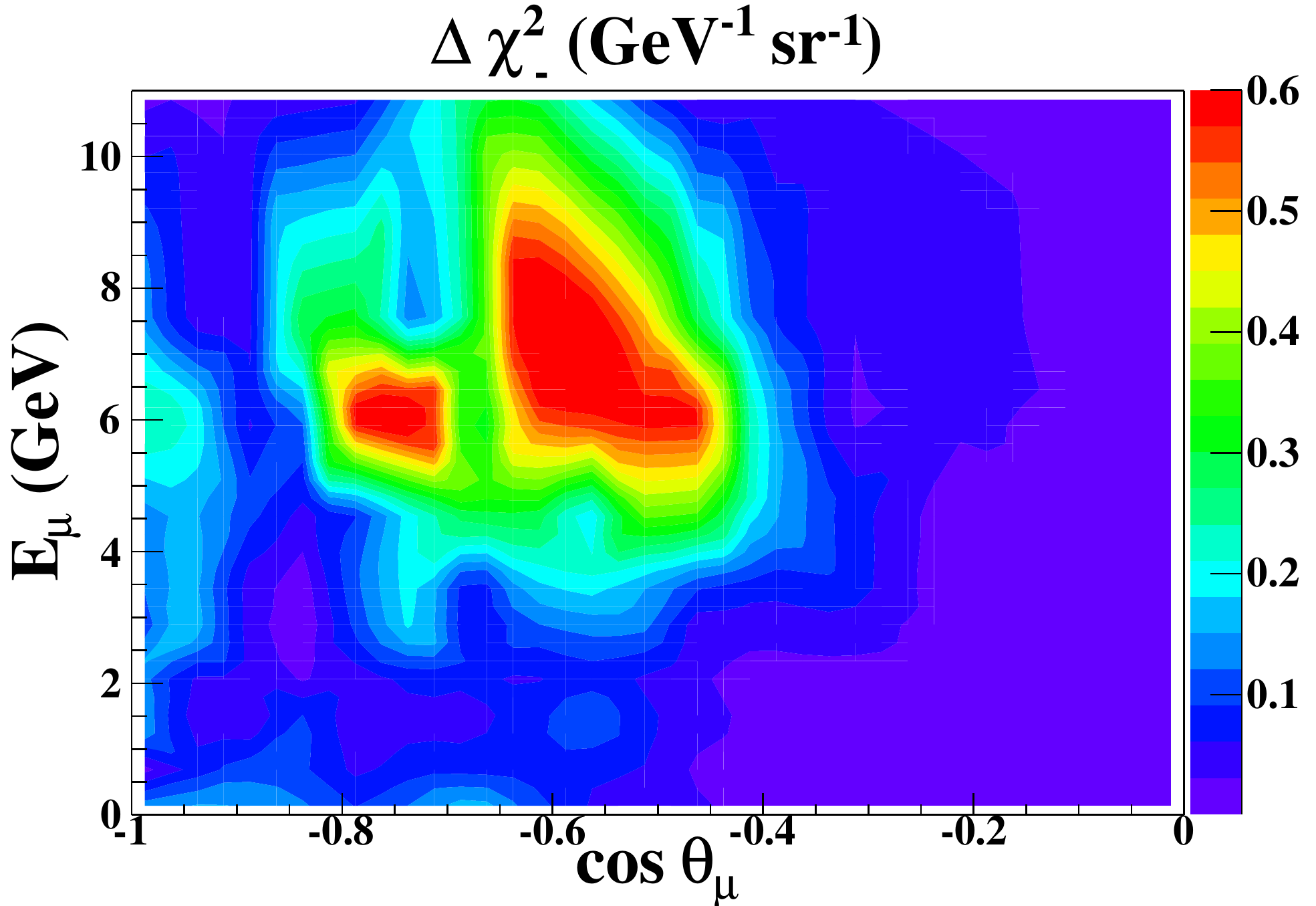}
\vspace{0.3cm}
\includegraphics[width=0.49\textwidth]{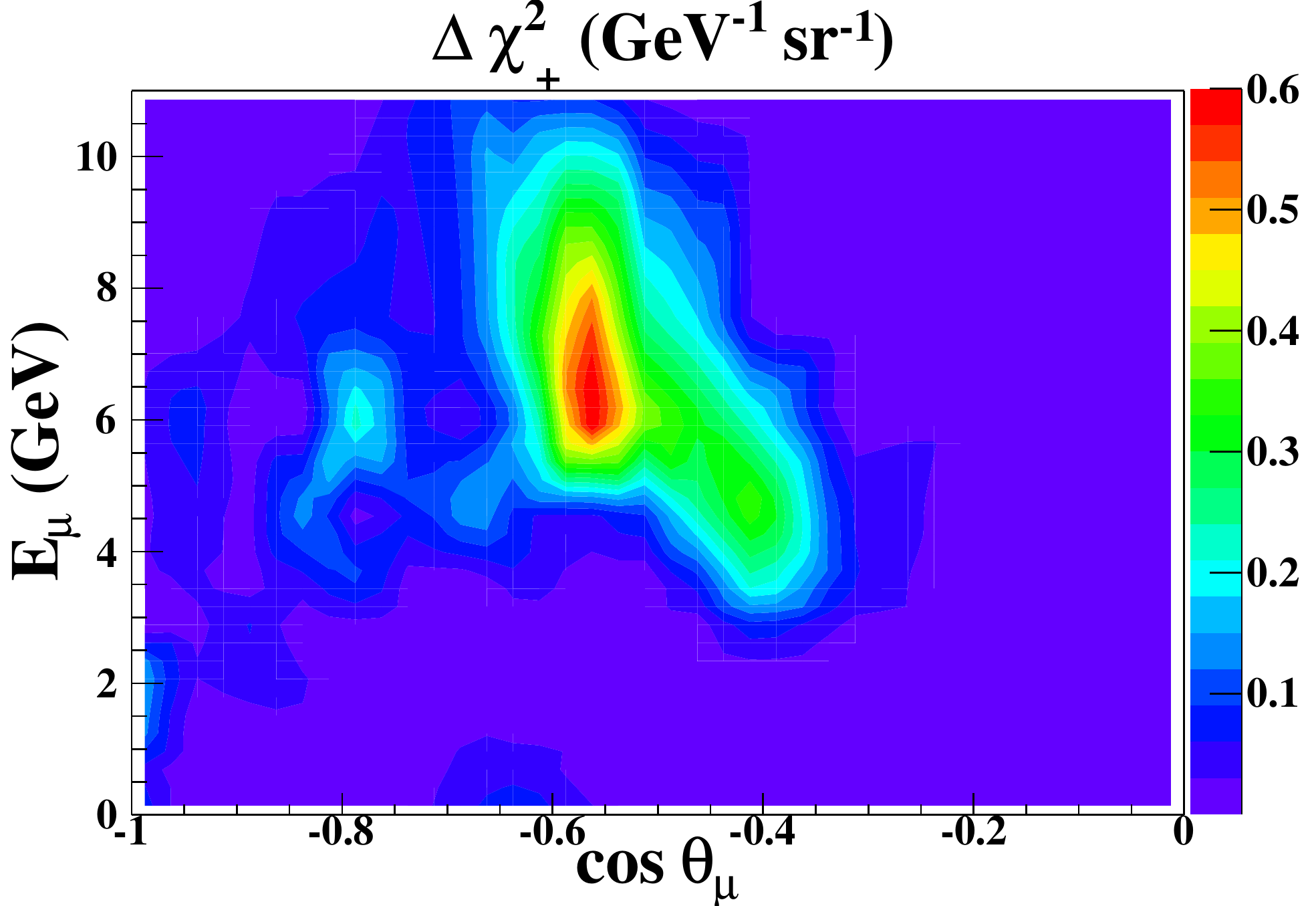}
\includegraphics[width=0.49\textwidth]{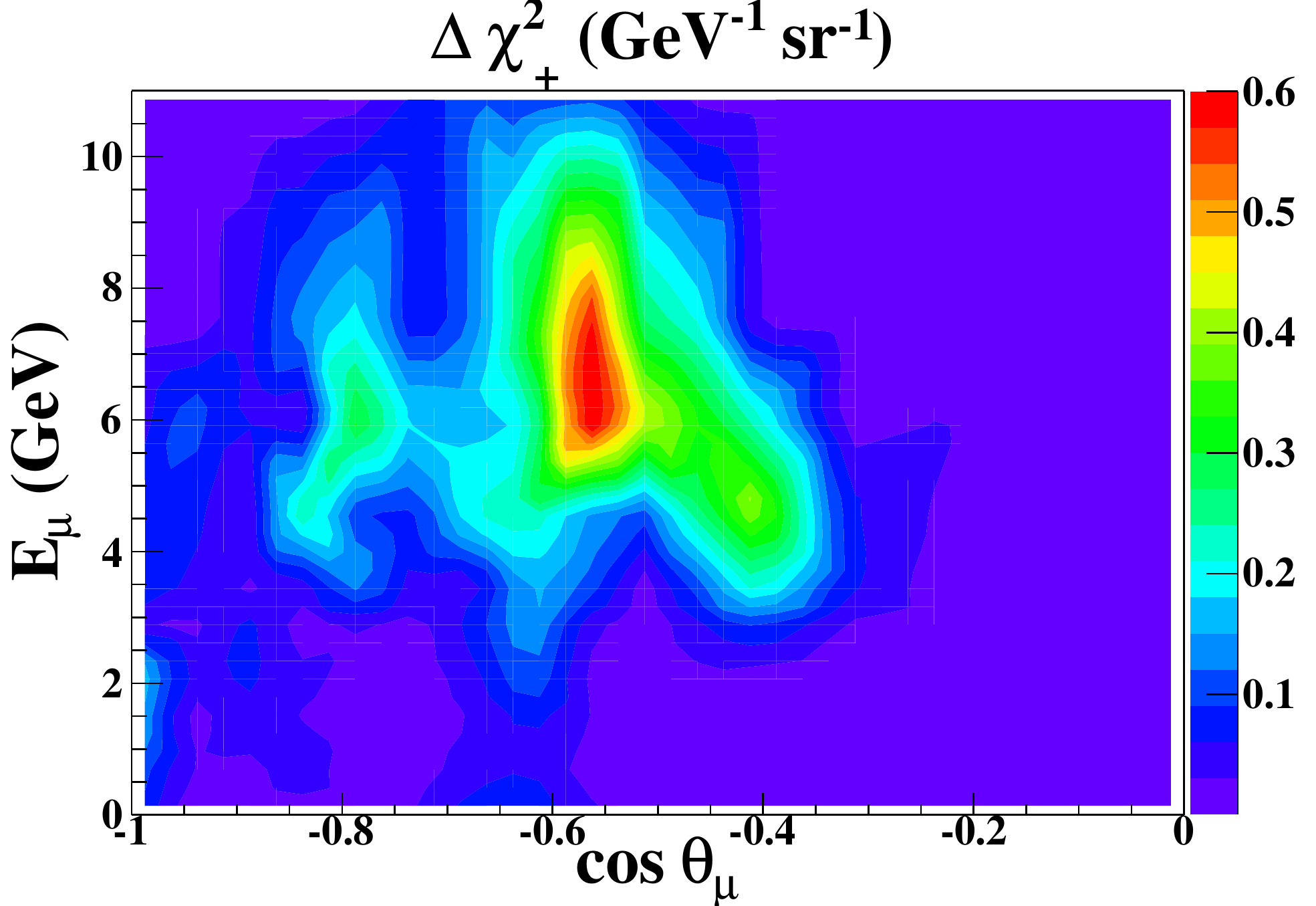}
\mycaption{Distribution of $\Delta \chi^2_-$ per unit area (top panels) 
and $\Delta \chi^2_+$ per unit area (bottom panels)  
in the $(E_\mu$--$\cos\theta_\mu)$ plane, when
NH is taken to be the true hierarchy.
The left panels show the distribution when hadron energy information
is not used. The right panels show the distribution when hadron energy 
information is used by further subdividing the events into four hadron 
energy bins. We take 500 kt-yr of ICAL exposure.}
\label{2d-3d-performance-comparison-new_mun}
\end{figure}

Before presenting the physics reach of ICAL for identifying the MH,
we motivate the extent to which the hadron energy information enhances the
capability of the experiment for this identification. 
In Fig.~\ref{2d-3d-performance-comparison-new_mun}, 
we show the distribution of $\Delta \chi^2_\pm \equiv \chi^2_\pm{\rm (IH)}-\chi^2_\pm{\rm (NH)}$ 
in the reconstructed $E_\mu$--$\cos\theta_\mu$ plane 
for the analysis that does not use the hadron energy information
(left panels) and the analysis where events are further divided into four
sub-bins depending on the reconstructed hadron energy (right panels). 
For the sake of this comparison, we do not consider the constant contribution in $\chi^{2}$ 
coming from the term involving the five pull parameters $\xi_{l}^{2}$ in Eq.~(\ref{chisq}). 
Also, we do not marginalize over the oscillation parameters in the fit. 
(For our final results, we do take care of the full pull contributions and marginalizations.)  

The upper (lower) panels in Fig.~\ref{2d-3d-performance-comparison-new_mun} 
depict the distribution of $\Delta \chi^2_{-}$ ($\Delta \chi^2_{+}$) coming from 
$\mu^{-}$ ($\mu^{+}$) events.
It can be observed that with the addition of the hadron energy 
information, the area in the $E_\mu$--$\cos\theta_\mu$ plane that
contributes significantly to $\Delta\chi^2_\pm$ increases, consequently
increasing the net $\Delta\chi^2_\pm$. Note that this increase in $\chi^2_\pm$ 
is not just due to the information contained in the hadron energy measurement,
but also due to that in the correlation between hadron energy
and muon momentum.

Another important point to be noted is that the increase in the sensitivity is not simply 
due to the events with low $E'_{\rm had}$, where the muon energy $E_{\mu}$ could be 
expected to closely match the original neutrino energy $E_{\nu}$. This may be seen 
from Table~\ref{tab_contribution_individual_hadron_bin}, where we present the total 
$\Delta \chi^2$ contributions from $\mu^{-}$ ($\mu^{+}$) events for the four individual hadron bins.
\begin{table}[h]
\begin{center}
\begin{tabular}{|c|l|l|l|}
\hline
$E'_{\rm had}$ (GeV) & events & $\Delta \chi^2$&$\Delta \chi^2$/events \\
 \hline
0 - 1 & 3995&5.8&0.0014\\
\hline
1 - 2 & 1152&1.9&0.0017\\
\hline
2 - 4 & 742&1.7&0.0023\\
\hline
4 -15 &677&1.2&0.0018\\
\hline
0 - 15 & 6566 & 10.7 & 0.0016\\
(with $E'_{\rm had}$ information) & & & \\
\hline
without $E'_{\rm had}$ information & 6775 & 6.3 & 0.0009 \\
\hline
\end{tabular}
\caption{Contributions of various $E'_{\rm had}$-bins to the total $\Delta\chi^2$. 
The events in the last row without $E'_{\rm had}$ information have true hadron 
energies up to 100 GeV. The same conditions as used for preparing 
Fig.~\ref{2d-3d-performance-comparison-new_mun} have been used here.} 
\label{tab_contribution_individual_hadron_bin}
\end{center}
\end{table}

Table~\ref{tab_contribution_individual_hadron_bin} shows that, while the 
$\Delta \chi^2$ contribution from the lowest $E'_{\rm had}$ bin is more than 
half the total $\Delta \chi^2$, this bin also has the majority of the total number of events.
Indeed, the normalized $\Delta \chi^2$ per event 
(see the last column of Table~\ref{tab_contribution_individual_hadron_bin})
is slightly higher for larger $E'_{\rm had}$ bins. This indicates that the hadron 
energy information from even the higher $E'_{\rm had}$ bins is significant 
for discriminating between the two mass orderings.

\begin{figure}[t]
\centering
\includegraphics[width=0.49\textwidth]{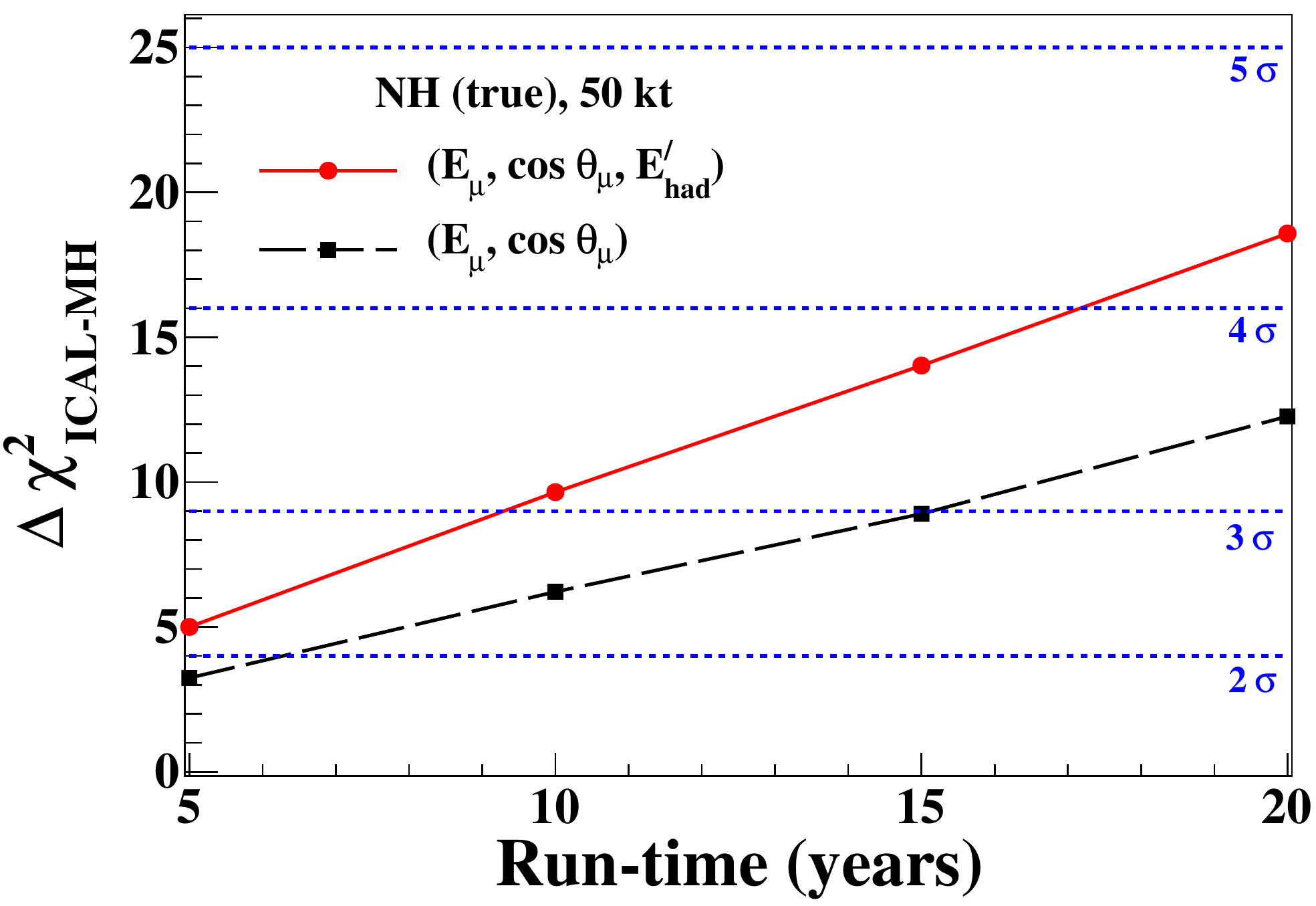}
\includegraphics[width=0.49\textwidth]{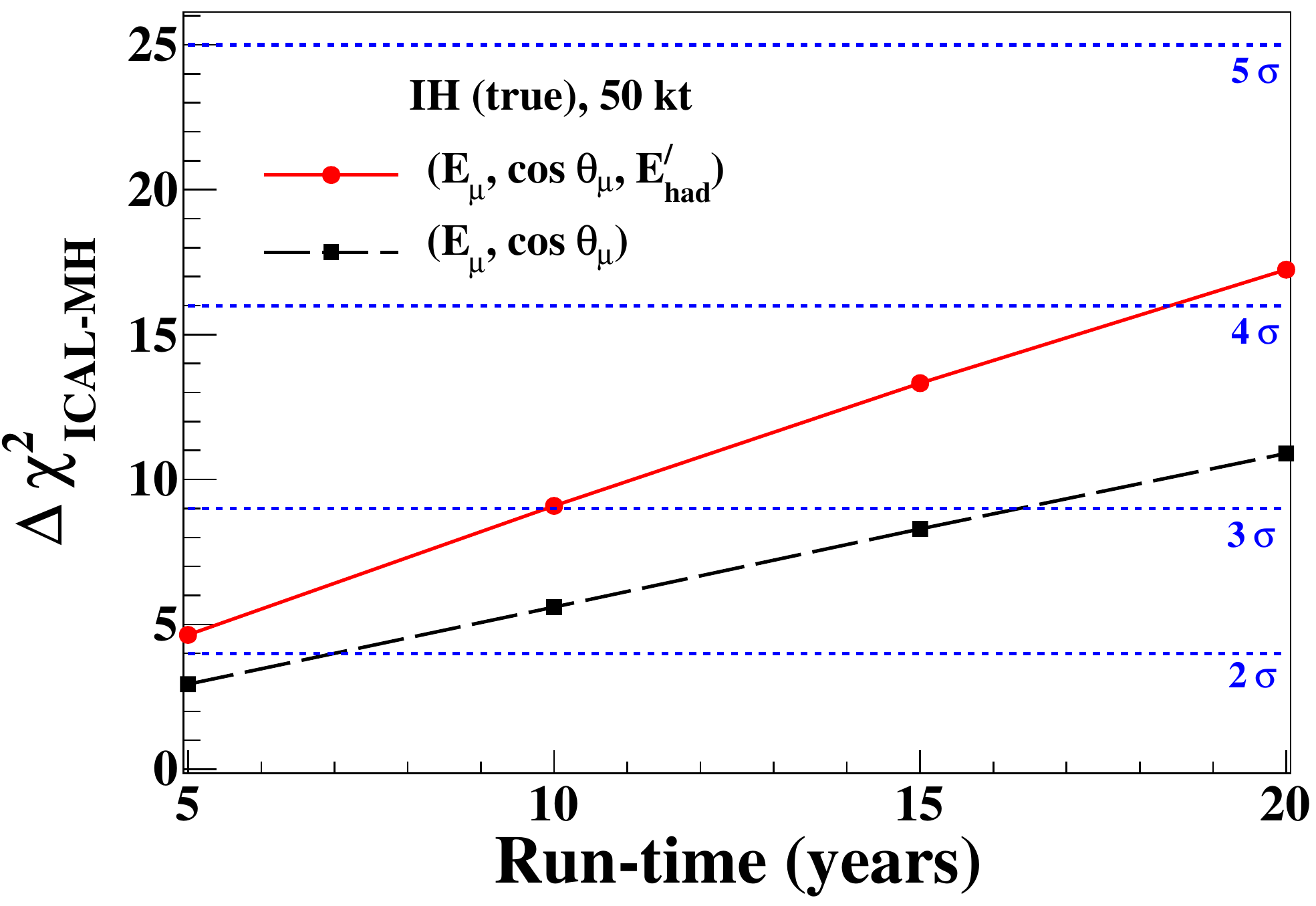}
\mycaption{$\chisqmh$ as a function of the run-time 
assuming NH (left panel) and IH (right panel) as true hierarchy.
The line labelled $(E_\mu, \cos\theta_\mu)$ denotes results without
including hadron information, while
the line labelled $(E_\mu, \cos\theta_\mu,E'_{\rm had})$ denotes 
improved results after including hadron energy information.
Here we have taken $\stch$(true) = 0.1 and $\sa$(true) = 0.5.}
\label{hierarchy}
\end{figure}

Figure~\ref{hierarchy} shows the sensitivity of 50 kt ICAL for 
identifying the neutrino mass hierarchy as a function of the run-time 
of the experiment. We find that after including the hadron energy 
information, 10 years of running can rule out the wrong hierarchy with 
$\chisqmh \approx 9.7$ (for true NH), and $\chisqmh \approx 9.1$ (for true IH).
Equivalently, the wrong hierarchy can be ruled out to 
about 3$\sigma$ for either hierarchy. 
This may be compared with the results without using hadron information.
The figure shows that for the same run-time, the value of 
$\chisqmh$ increases by about 40\% when the correlated
hadron energy information is added.
Note that for the comparison here, we have used the same binning scheme
in $(E_\mu,\cos\theta_\mu)$ as shown in Table~\ref{table:3d-bin} for both 
analyses. One may use finer binning for the analysis without hadron 
information, as has been done in \cite{Ghosh:2012px}. 
We find that the $\chisqmh$ with hadron energy information added
is about 35\% more than that in \cite{Ghosh:2012px}.
Thus, the improvement seen in our analysis with hadron energy information is 
not merely due to using additional bins compared to the muon-only analysis.
Here we would like to point out that, in ICAL, we can explore the Earth's 
matter effect in neutrino and antineutrino channels {\it separately} using its charge 
identification capability via magnetic field. This feature is very crucial in order
to enhance the sensitivity to MH. 

\begin{figure}[t]
\centering
\includegraphics[width=0.49\textwidth]{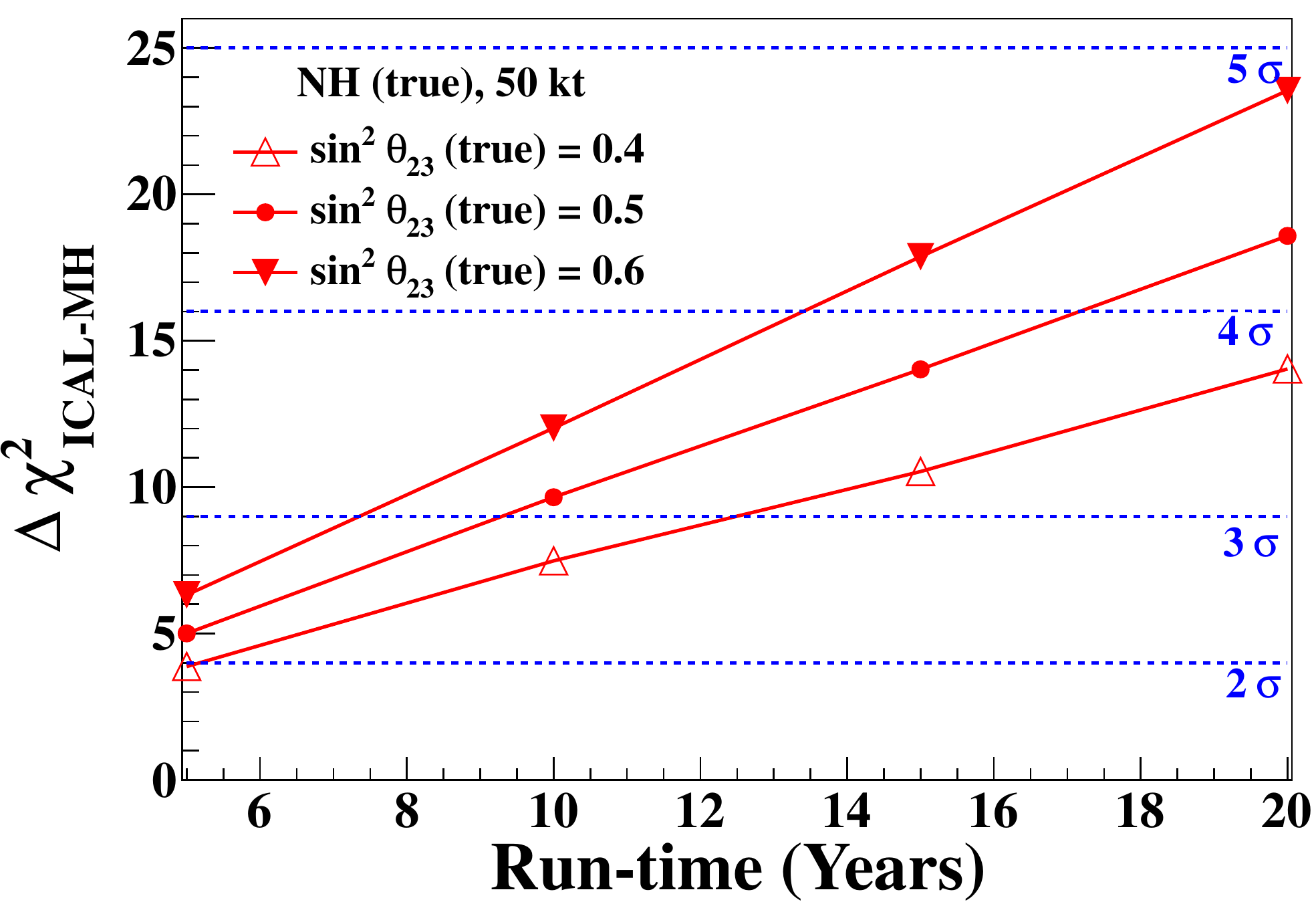}
\includegraphics[width=0.49\textwidth]{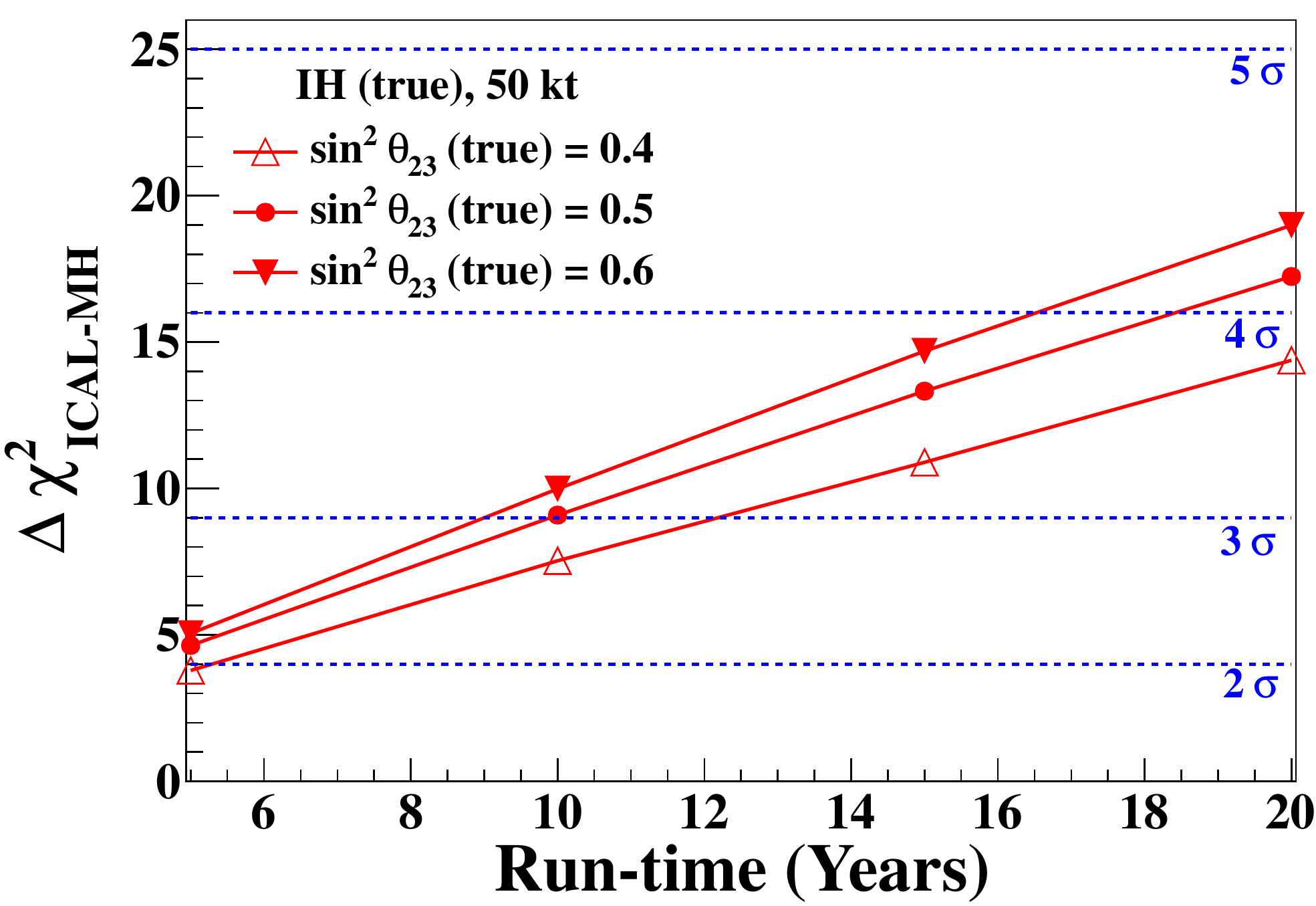}
\mycaption{Variation of $\chisqmh$ 
for different true values of $\sa$. The left panel (right panel) 
shows the results assuming NH (IH) as true hierarchy.
Here we have taken $\sin^2 2\theta_{13}{\rm (true)} = 0.1$.} 
\label{hierarchy1}
\end{figure}

\begin{figure}[t]
\centering
\includegraphics[width=0.49\textwidth]{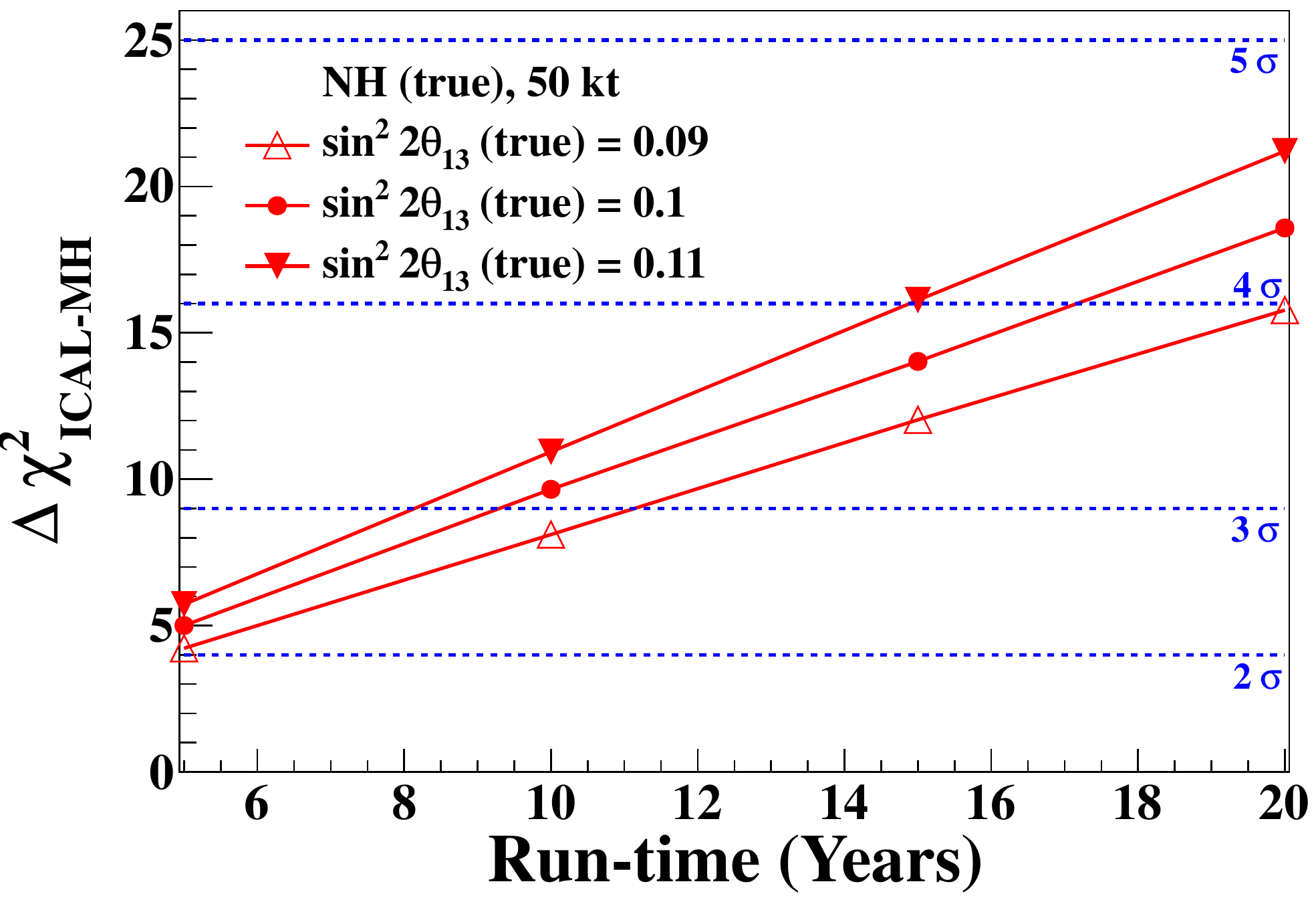}
\includegraphics[width=0.49\textwidth]{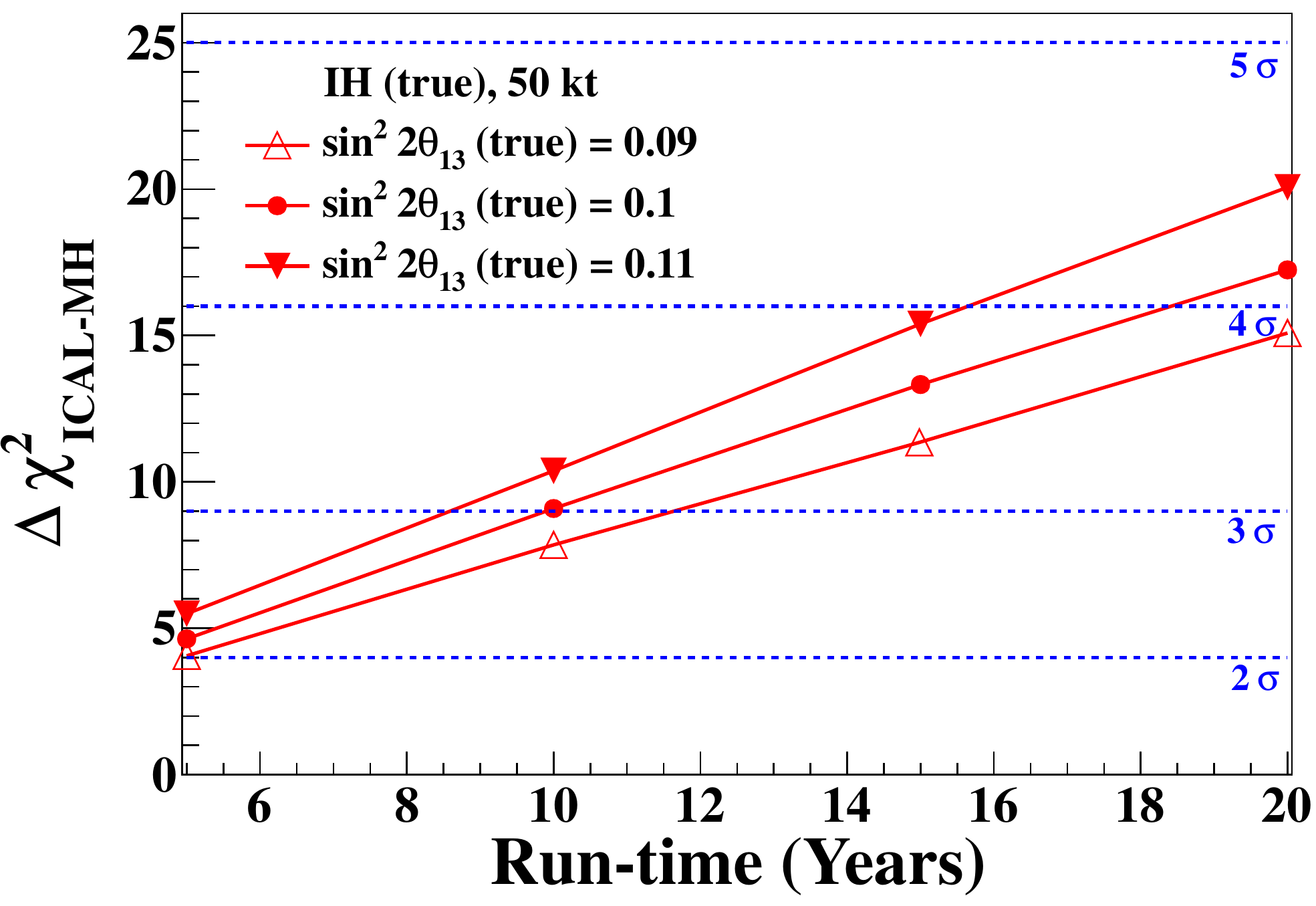}
\mycaption{Variation of $\chisqmh$ 
for different true values of $\stch$. The left panel (right panel) 
shows the results assuming NH (IH) as true hierarchy.
Here we have taken $\sin^2 \theta_{23}{\rm (true)} = 0.5$.} 
\label{hierarchy2}
\end{figure}

Figures~\ref{hierarchy1} and \ref{hierarchy2} show the variation of the MH
identification potential for three benchmark values of $\sa$ and $\stch$,
respectively, in the allowed ranges of these parameters.
Higher values of $\sa$ and $\stch$ increase the 
matter effects in neutrino oscillations and thus result in better 
hierarchy sensitivity, as seen in these plots. 
This is expected since the leading matter effect terms in the probability 
expressions of $P_{\mu\mu}$ and $P_{e\mu}$
are proportional to these parameters \cite{Akhmedov:2004ny}.
Depending on the true values of these parameters and the true choice of MH, 
the ICAL detector can identify the MH with a $\chisqmh$ in the range of 
$7$ to $12$ using an exposure of 500 kt-yr.

As for the variation of $\chisqmh$ with respect to $\dcp$,
we have checked that the projected ICAL atmospheric data 
is not sensitive to $\dcp$, even after the addition of hadron energy information. 
This is not surprising considering the fact that in the expression of $P_{\mu\mu}$, 
the $\dcp$ dependent term is suppressed by a factor of 
$\alpha \equiv \Delta m^2_{21}/\Delta m^2_{31}$ \cite{Akhmedov:2004ny}. 
Recently, this feature of the atmospheric data has also been verified in
\cite{Blennow:2012gj,Ghosh:2012px}.

\subsection{Precision Measurement of Atmospheric Parameters}
\label{PM-results}

In order to quantify the precision in the measurements of a parameter 
$\lambda$ (here $\lambda$ may be $\sin^2\theta_{23}$ or $|\Delta m^2_{32}|$ or both), 
we use the quantity:
\begin{equation}
\chisqpm(\lambda) = 
\chisqical(\lambda) - \chi^2_0 \; ,
\end{equation}
where $\chi^2_0$ is the minimum value of $\chisqical$ in the
allowed parameter range. 
Here with the statistical fluctuations suppressed, $\chi^2_0 \approx 0$.
The significance is denoted by $n\sigma$ 
where $n \equiv \sqrt{\chisqpm}$. 
In terms of these quantities, we define the relative precision achieved on the parameter 
$\lambda$ at $1\sigma$ as \cite{Fogli:2012ua}
\begin{equation}
p(\lambda) = \frac{\lambda {\rm (max)} - \lambda{\rm (min)}}
{4 ~\lambda{\rm (true)}},
\label{pm_def}
\end{equation}
where $\lambda$(max) and $\lambda$(min) are the maximum and minimum allowed values 
of $\lambda$ at $2\sigma$ respectively, and $\lambda{\rm (true)}$ is its true choice.

\begin{figure}[t]
\centering
\includegraphics[width=0.49\textwidth]{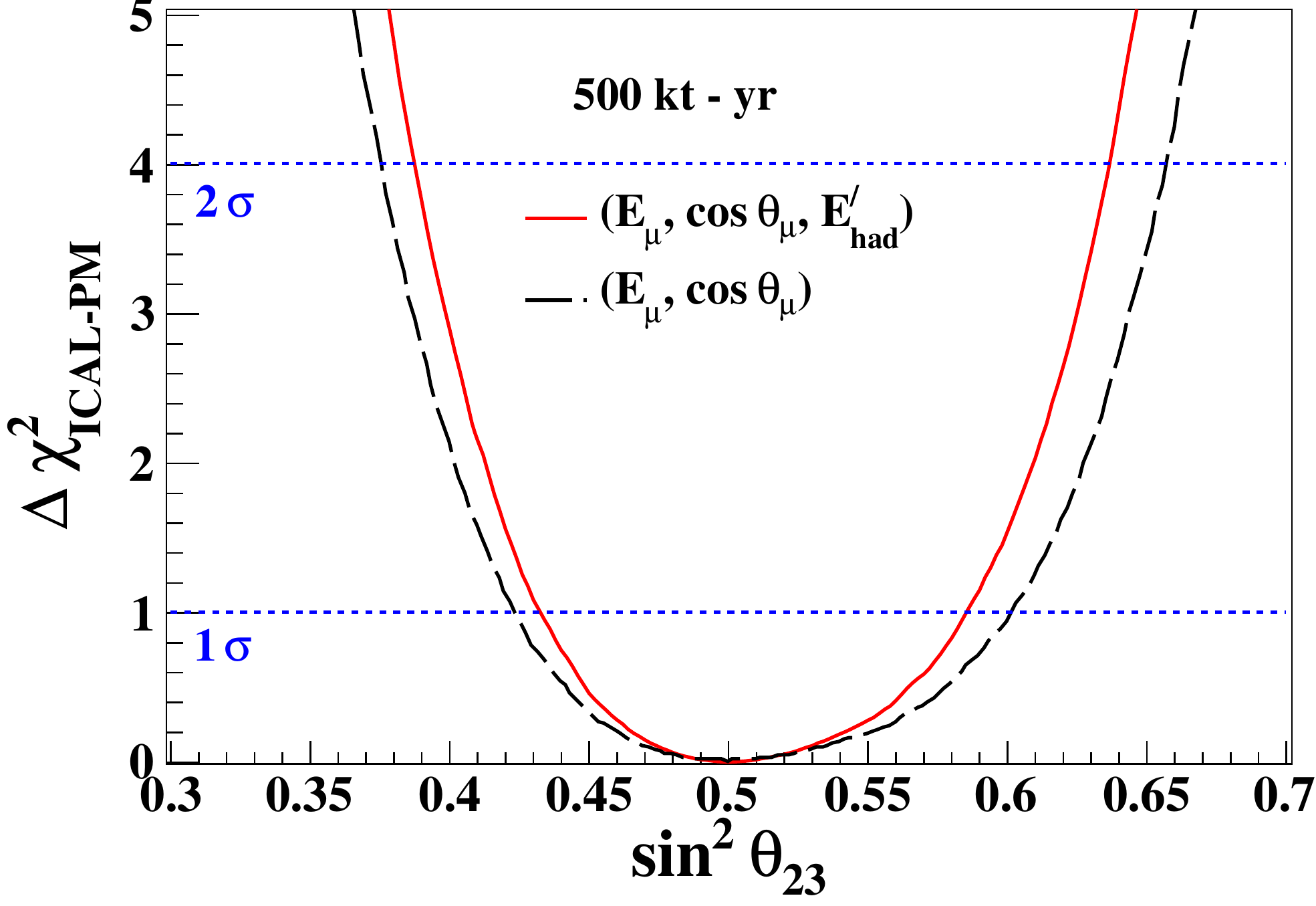}
\includegraphics[width=0.49\textwidth]{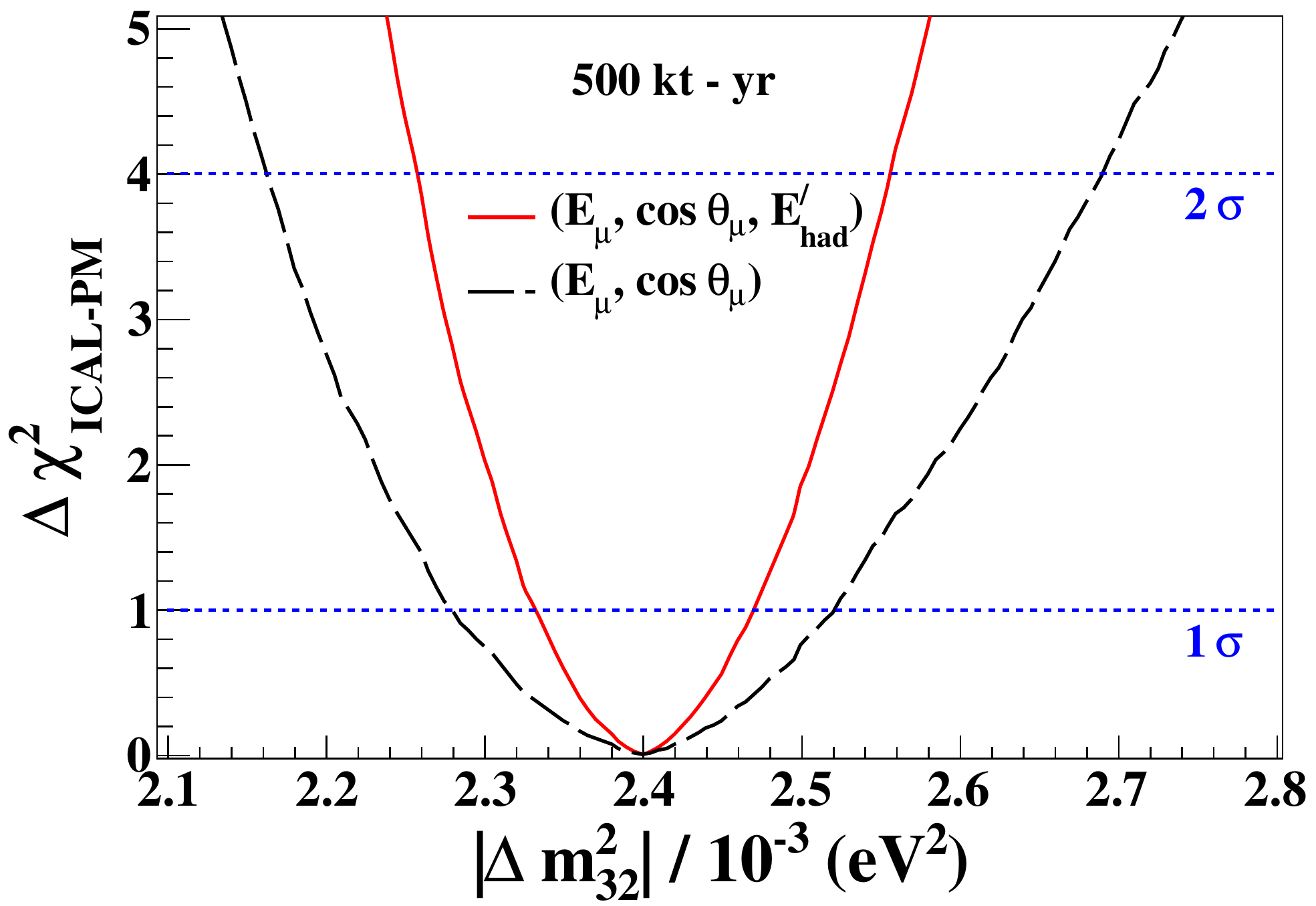}
\mycaption{The left panel shows  $\chisqpm(\sa)$ and 
the right panel depicts $\chisqpm(|\mam|)$, assuming
NH as true hierarchy. The lines labelled $(E_\mu, \cos\theta_\mu)$ 
denote results without including hadron information, while
the lines labelled $(E_\mu, \cos\theta_\mu,E'_{\rm had})$ denote 
improved results after including hadron energy information.}
\label{pm-1d}
\end{figure}

In the two panels of Fig.~\ref{pm-1d}, we show the sensitivity of ICAL
to the two parameters $\sa$ and $|\mam|$ separately, where the other 
parameter has been marginalized over. We also marginalize over 
$\theta_{13}$ and the two possible choices of mass hierarchies.
While the figure shows the results for NH as the true hierarchy,
we have checked that the results with true IH are almost identical.
It may be observed from the figure that with the inclusion of hadron 
energy information, 500 kt-yr of ICAL exposure would be able to measure  
$\sa$ to a relative $1\sigma$ precision of 12\% and $|\Delta m^2_{32}|$ to 2.9\%. 
With the muon-only analysis, the same relative precisions would be
13.7\% and 5.4\%, respectively.
Note that, for this comparison, the binning for $(E_\mu, \cos\theta_\mu)$ 
has been kept identical in both scenarios, with and without 
hadron information.
One may use finer binning for the analysis without hadron information.
However, we have checked that even the muon-only analysis with finer binning 
(20 $E_\mu$ bins and 80 $\cos\theta_\mu$ bins)
can yield a precision only up to 13.5\% and 4.2\%, respectively,
for the $\sa$ and $|\mam|$ precision.

The observations above may be understood by noting that the $\sa$ precision 
is governed mainly by the statistics available to the experiment, which
does not change by adding the hadron energy information, and therefore the
addition of hadron energy information  makes only a small difference in the 
two analyses. 
On the other hand, independent measurements of $E_\mu$ and $E'_{\rm had}$
corresponds to a better estimation of $E_\nu$, which appears in the
oscillation expression as $\sin^2(\Delta m^2 L/E_\nu)$. A better
measurement of $E_\nu$ thus leads to a better measurement of 
$\Delta m^2$, resulting in the dramatic improvement in the 
precision on $\mam$ observed here.

\begin{figure}[t]
\centering
\includegraphics[width=0.49\textwidth]{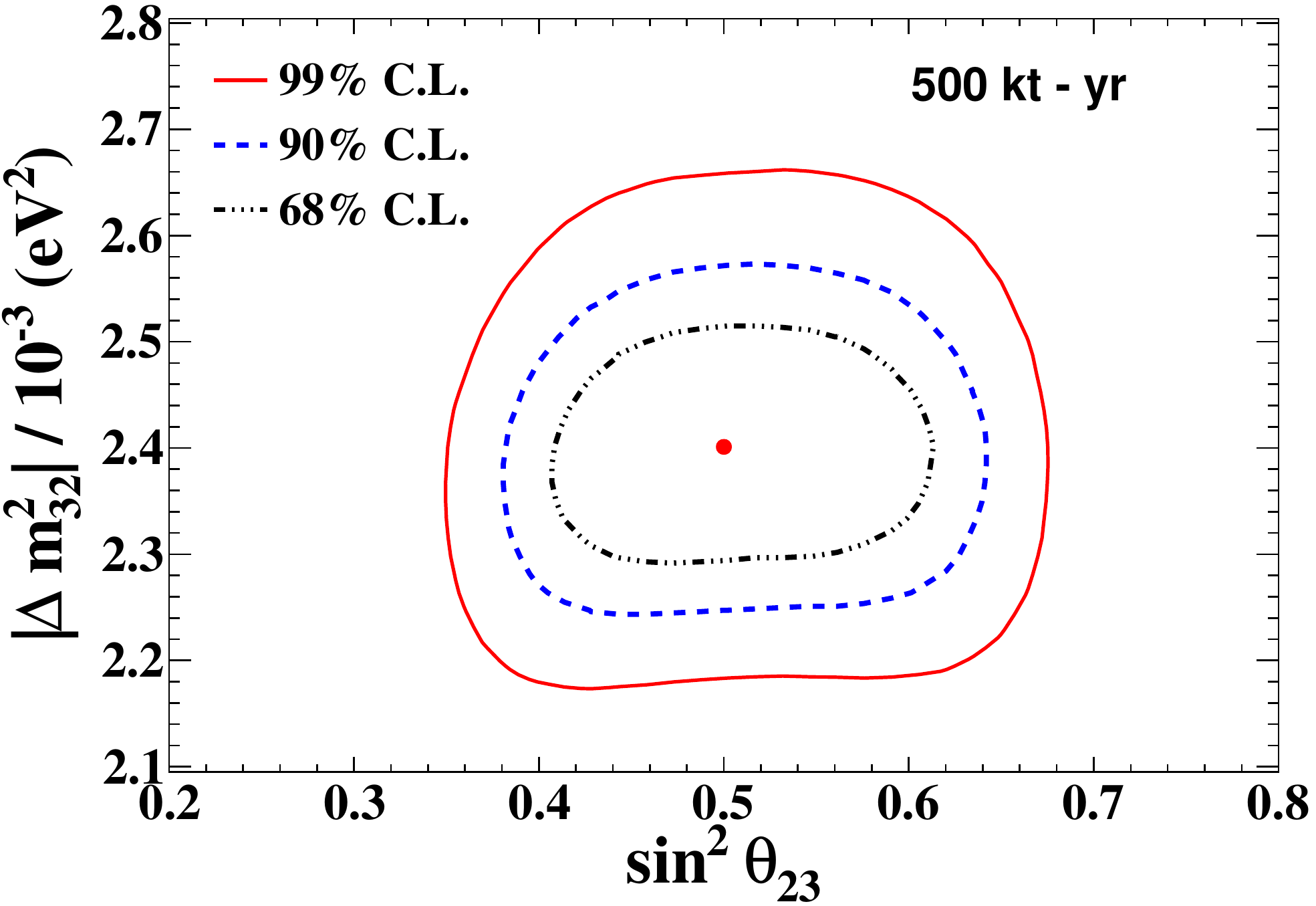}
\includegraphics[width=0.49\textwidth]{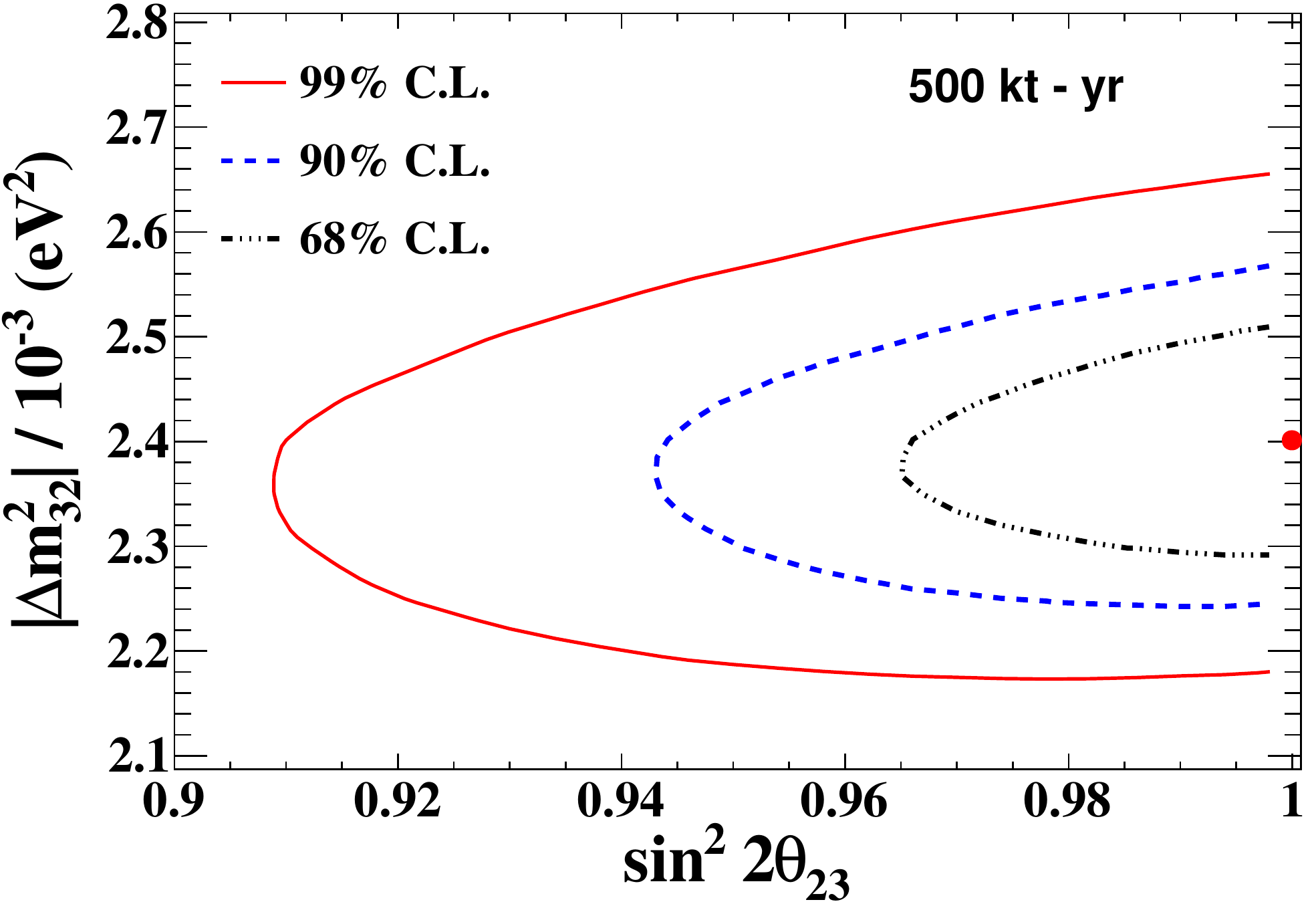}
\mycaption{$\chisqpm$ contours at 68\%, 90\%, and 99\%
confidence levels (2 dof) in $\sa$--$|\mam|$ plane (left panel) 
and in $\sta$--$|\Delta m^2_{32}|$ plane (right panel)
after including the hadron energy information. We assume NH as the true hierarchy.
The true choices of the parameters have been marked with a dot.} 
\label{pm-ssqth-ssq2th-mmix}
\end{figure}

\begin{figure}[t]
\centering
\includegraphics[width=0.75\textwidth]{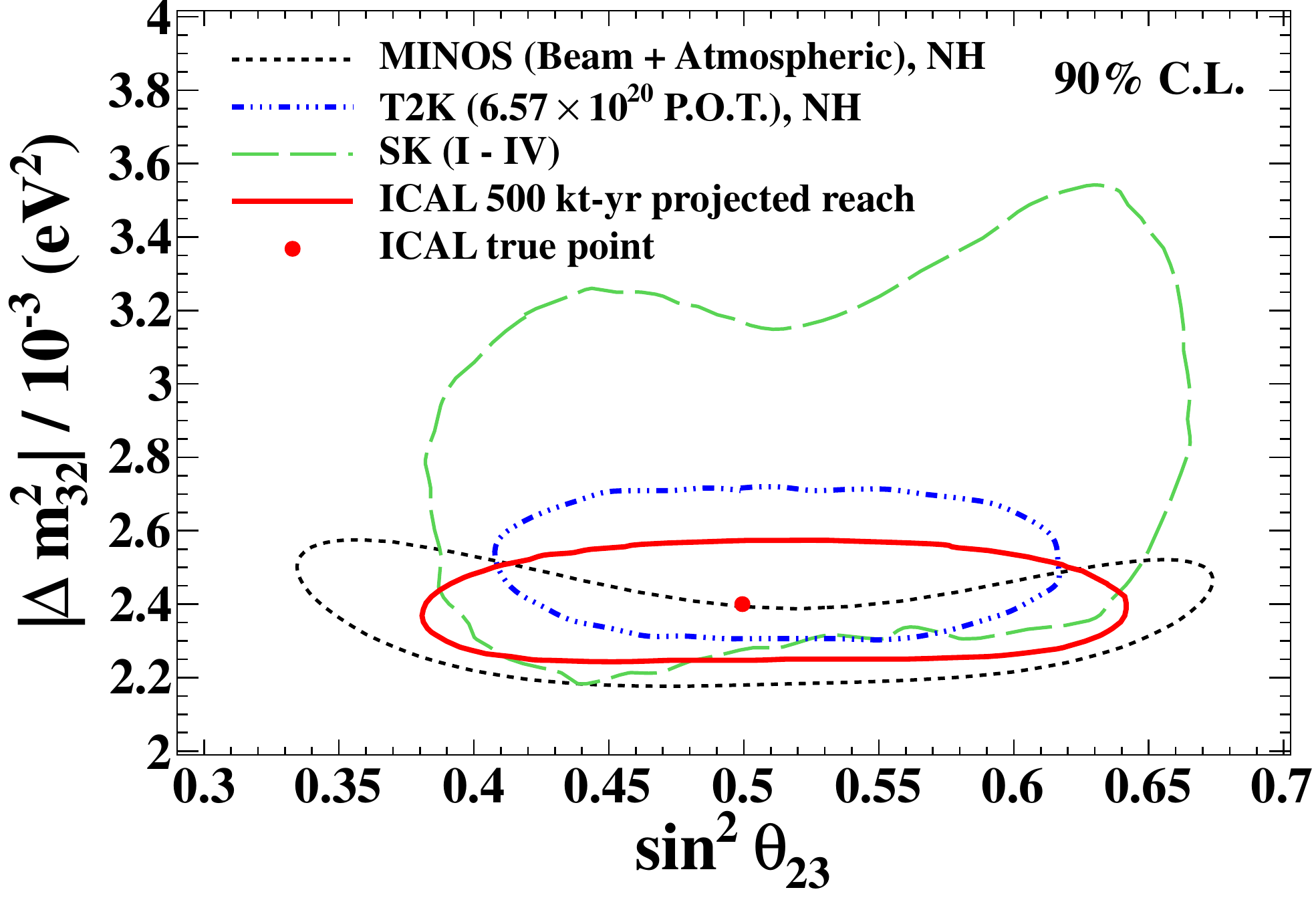}
\mycaption{90\% C.L. (2 dof) contours in the $\sa$--$|\Delta m^2_{32}|$ plane.
The current limits from Super-Kamiokande \cite{Himmel:2013jva},
MINOS \cite{Adamson:2014vgd}, and T2K \cite{Abe:2014ugx}
have been shown along with the projected ICAL reach for the exposure of
500 kt-yr, assuming true NH. The true choices of the parameters for ICAL 
have been marked with a dot.} 
\label{3d-pm-ssq2th-mmix}
\end{figure}

Figure~\ref{pm-ssqth-ssq2th-mmix} shows the $\chisqpm$ 
contours at 68\%, 90\%, and 99\% confidence levels in the 
$\sa$--$|\mam|$ plane (left panel) and in the $\sta$--$|\Delta m^2_{32}|$ 
plane (right panel) after including the hadron energy information.
Here the true value of $\theta_{23}$ has been taken to be maximal,
so the contours in the left panel are almost symmetric in $\sa$. 
The comparison of the projected 90\% C.L. precision reach of ICAL 
(500 kt-yr exposure) in $\sa$--$|\Delta m^2_{32}|$ plane
with other experiments \cite{Himmel:2013jva,Adamson:2014vgd,Abe:2014ugx}
is shown in Fig.~\ref{3d-pm-ssq2th-mmix}. 
Using hadron energy information, the ICAL will be able to achieve a
$\sa$ precision comparable to the current precision for Super-Kamiokande 
\cite{Himmel:2013jva} or T2K \cite{Abe:2014ugx},
and the $|\mam|$ precision comparable to the MINOS reach 
\cite{Adamson:2014vgd}. 
Of course, some of these experiments would have collected much more 
statistics by the time ICAL would have an exposure of 500 kt-yr.
The ICAL will therefore not be competing with these experiments 
for the precision measurements of these mixing parameters, however
the ICAL measurements will serve as complementary information 
for the global fit of world neutrino data. 
Note that, as compared to the atmospheric neutrino analysis at
Super-Kamiokande, the ICAL precision on $|\mam|$ is far superior.
This is a consequence of the better precision in the reconstruction
of  muon energy and direction at ICAL.

\begin{figure}[t]
\centering
\includegraphics[width=0.49\textwidth]{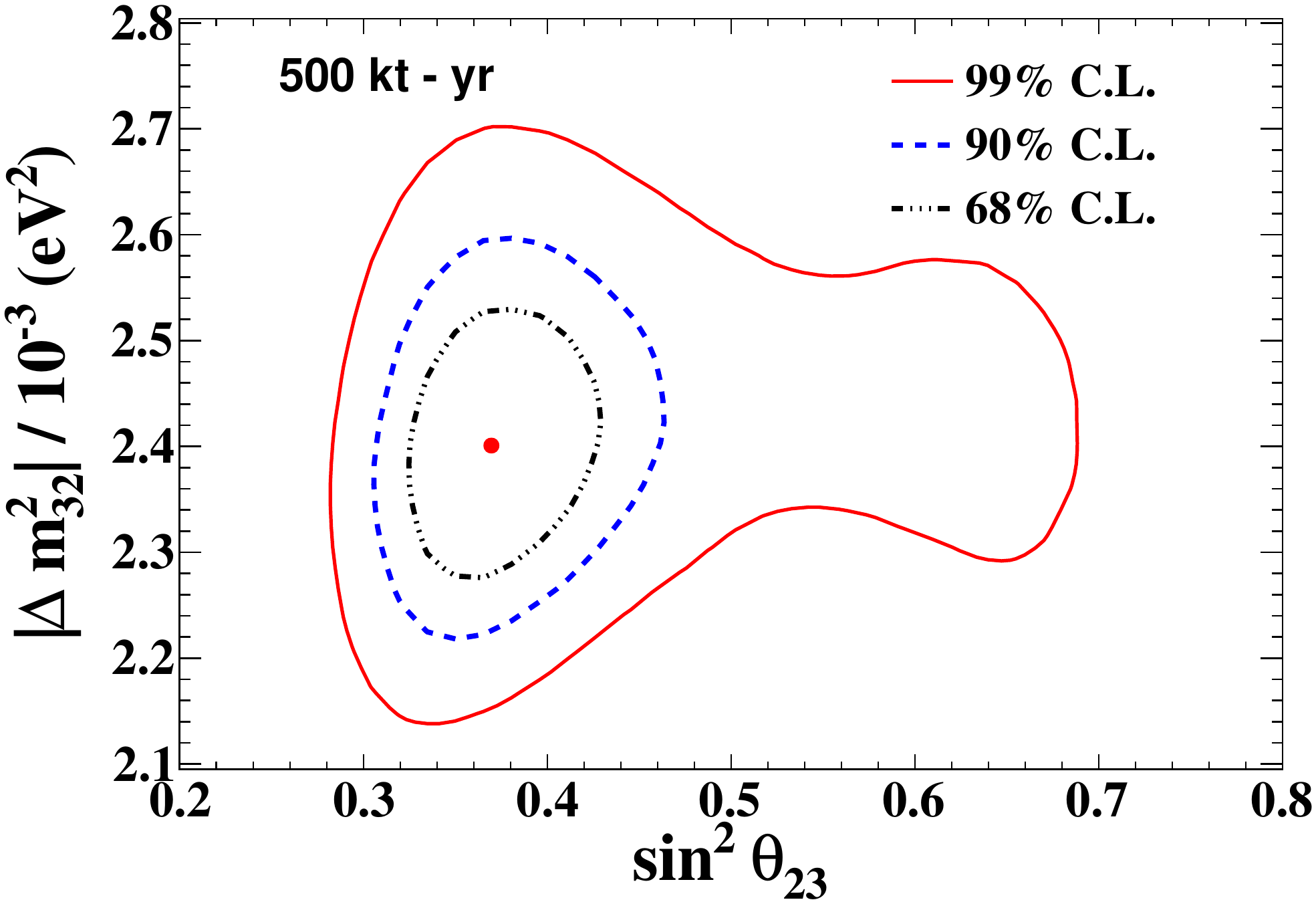}
\includegraphics[width=0.49\textwidth]{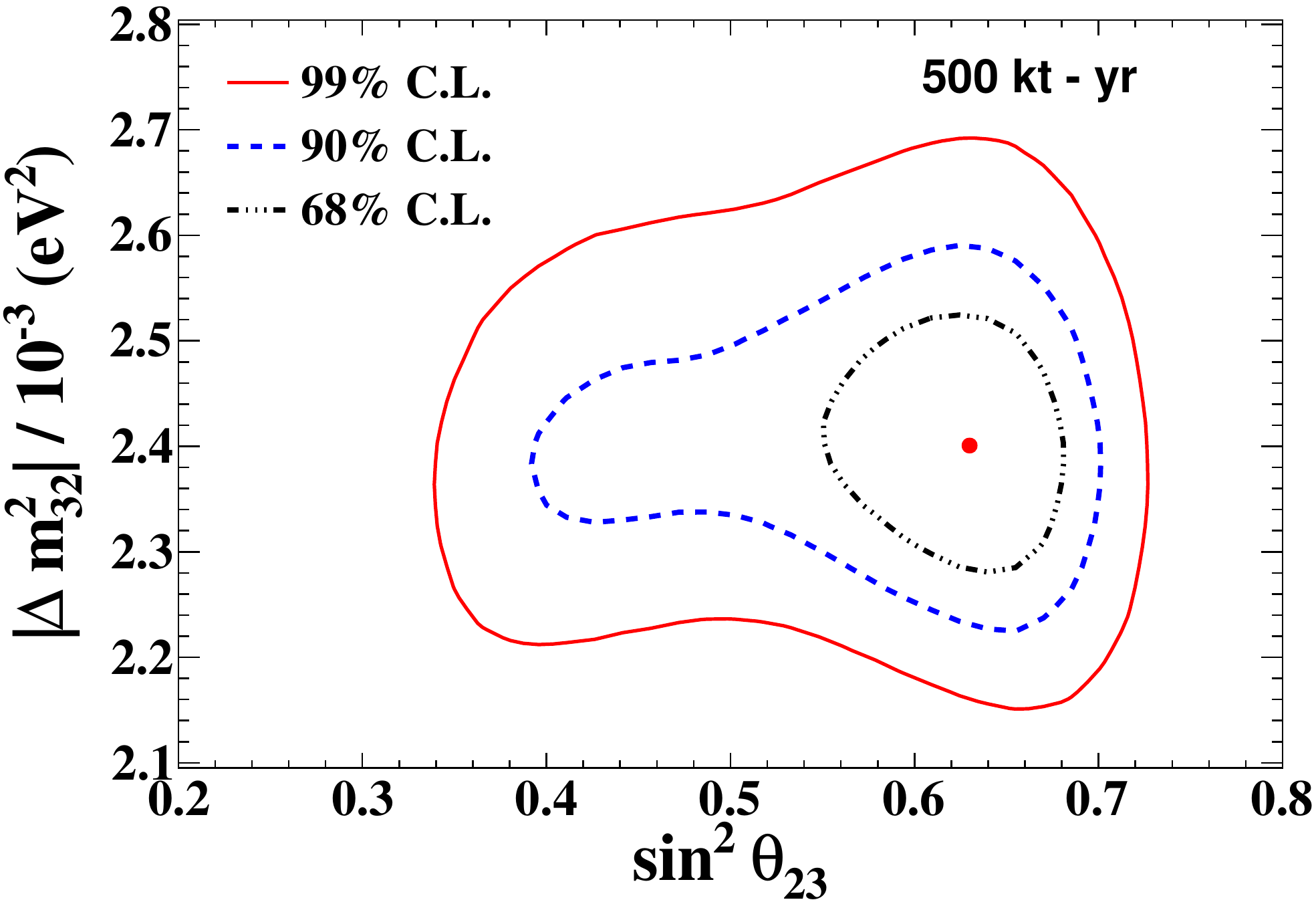}
\mycaption{$\chisqpm$ contours at 68\%, 90\%, and 99\%
confidence levels (2 dof) in $\sa$--$|\mam|$ plane,
for $\sa$(true) = 0.37 (left panel) and $\sa$(true)= 0.63 (right panel)
after including the hadron energy information. 
We assume NH as the true hierarchy.
The true choices of the parameters have been marked with a dot.} 
\label{PM-nonmaximal}
\end{figure}

Finally in this subsection, we present 68\%, 90\%, and 99\% C.L. contours 
in the $\sa$--$|\Delta m^2_{32}|$ plane, considering non-maximal
values of the mixing angle $\theta_{23}$. 
Figure~\ref{PM-nonmaximal} shows the sensitivity of ICAL for 
$\sta$ = 0.93 (i.e. $\sa$ = 0.37, 0.63). 
It can be seen that the precisions obtained are similar, though
the shapes of the contours are more complicated.
We observe that for $\theta_{23}$ in the lower octant, 
the maximal mixing can be ruled out with 99\% C.L. 
with 500 kt-yr of ICAL data. However, if $\theta_{23}$ is closer to the 
maximal mixing value, or in the higher octant, then the ICAL sensitivity 
to exclude maximal mixing would be much smaller. These contours can also 
be seen as precursors to resolving the $\theta_{23}$ octant degeneracy, 
which will be discussed in Sec.~\ref{OS-results}.

\subsection{Octant of $\tmt$}
\label{OS-results}

One can exploit the Earth's matter effect in the $P_{\mu\mu}$ channel to resolve the octant 
ambiguity of $\theta_{23}$ \cite{Choubey:2005zy}. In analogy with the mass hierarchy discovery 
potential, we quantify the statistical significance of the analysis to rule out the wrong octant by
\begin{equation}
\chisqos = \chisqical(\text{false octant}) - 
\chisqical (\text{true octant}) .
\label{os_chi2_def}
\end{equation}
Here $\chisqical(\text{true octant})$ 
and $\chisqical(\text{false octant})$
are obtained by performing a fit to the ``observed'' data assuming
the true octant and wrong octant, respectively. 
Here with the statistical fluctuations suppressed,
$\chisqical (\text{true octant}) \approx 0$.
For each given value of $\theta_{23}$(true), we marginalize over all
the allowed values of $\theta_{23}$ in the opposite octant, including the 
maximal mixing value. We also marginalize $\chisqos$ over the 
true choices of mass hierarchy in addition to the oscillation parameters 
mentioned in \ref{sec:numerical}. The statistical significance for ruling out 
the wrong octant is represented in terms of 
$n\sigma$, where $n \equiv \sqrt{\chisqos}$.

\begin{figure}[t]
\centering
\includegraphics[width=0.49\textwidth]{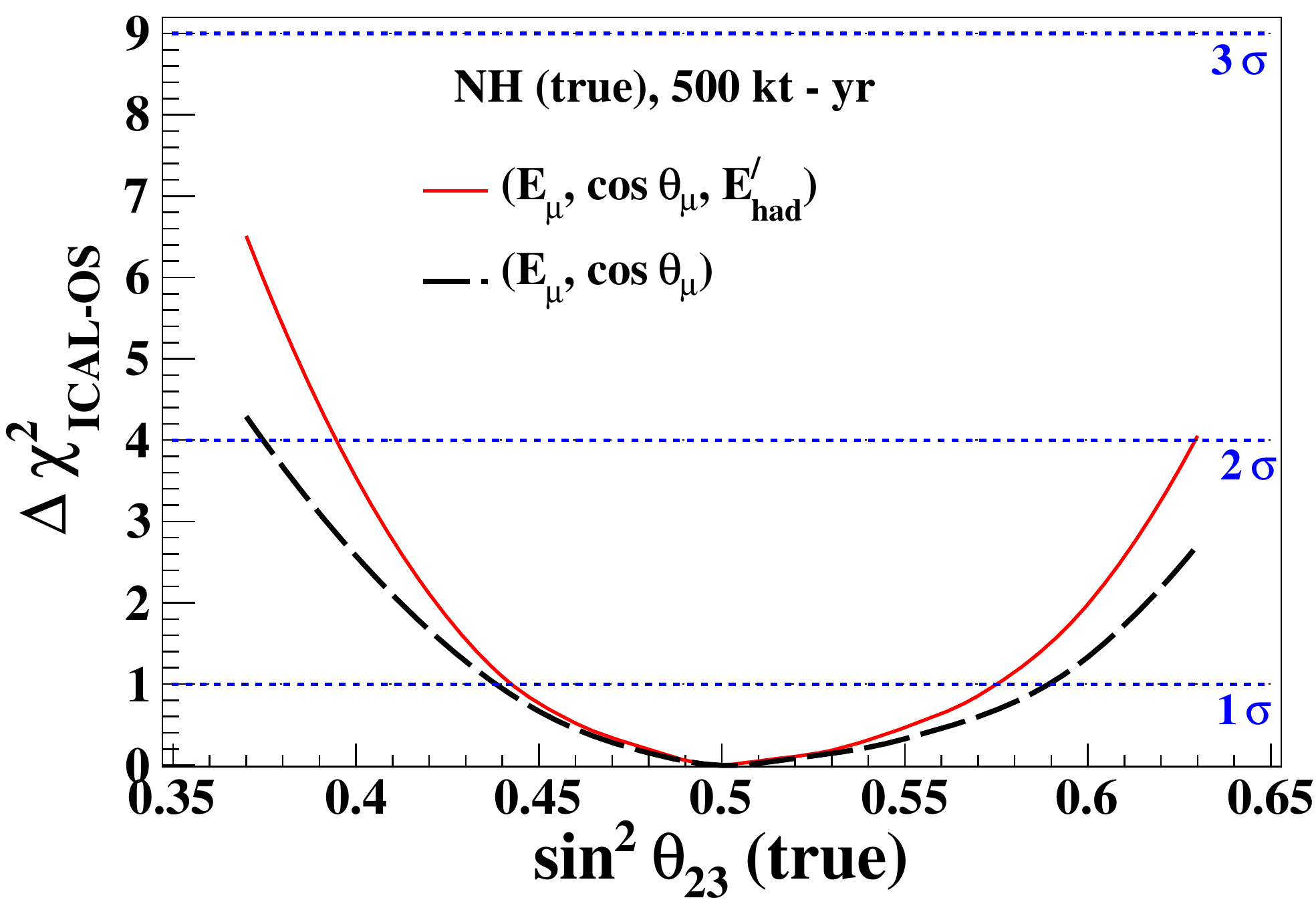}
\includegraphics[width=0.49\textwidth]{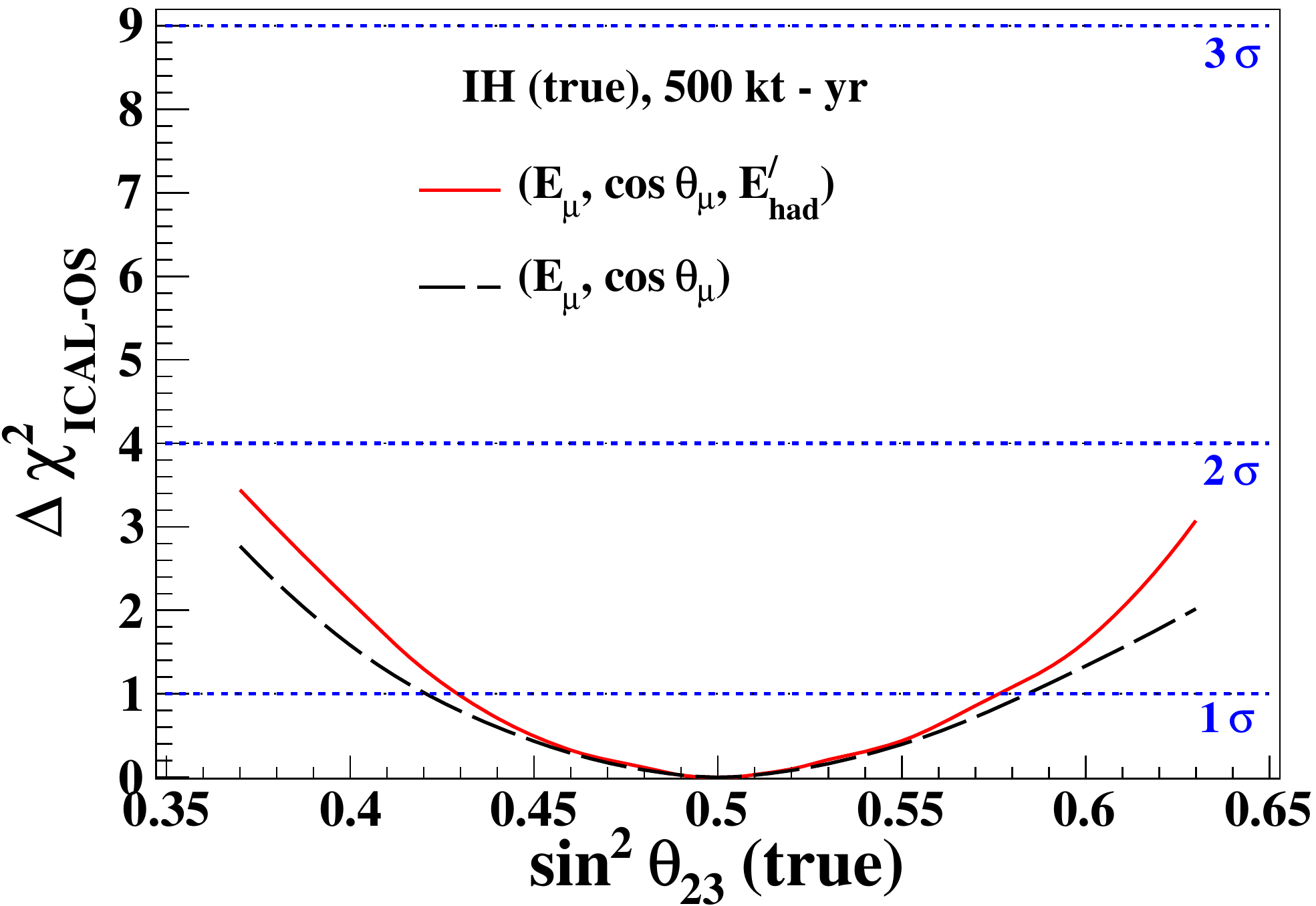}
\mycaption{$\chisqos$ 
for octant discovery potential as a function of true $\sa$.
The left panel (right panel) assumes NH (IH) as true hierarchy.
The line labelled $(E_\mu, \cos\theta_\mu)$ denotes results without
including hadron information, while
the line labelled $(E_\mu, \cos\theta_\mu,E'_{\rm had})$ denotes 
improved results after including hadron energy information.
ICAL exposure of 500 kt-yr is considered.}
\label{os-2d-vs-3d-th13-best-fit}
\end{figure}

Figure~\ref{os-2d-vs-3d-th13-best-fit} shows the sensitivity of ICAL 
to the identification of the $\theta_{23}$ octant,
with and without including the hadron energy information.
It may be observed that a $2\sigma$ identification of the octant
is possible with the 500 kt-yr INO data alone only when the true hierarchy 
is NH and the true octant is LO. In this case, without using the 
hadron energy information one can get a $2\sigma$ identification
only when $\sa (\text{true)}<0.375$, which is almost close to the present
$3\sigma$ bound. With the addition of hadron energy information, this
task is possible as long as $\sa (\text{true)}<0.395$. 
If the true octant is HO or the true mass hierarchy is inverted,
then the discrimination of $\theta_{23}$ octant with the ICAL data alone 
becomes rather difficult. In case of NH (IH), neutrino (antineutrino) events 
are mostly affected by the Earth's matter effect and give vital clues towards 
the octant of $\theta_{23}$. Since the statistical strength of atmospheric 
neutrino events is higher compared to antineutrino events, the octant sensitivity 
is better for NH compared to IH.
We have checked that these observations are not much sensitive to the 
true value of $\theta_{13}$. A variation of $\stch$(true) in the range 
$0.09$ to $0.11$ changes the values of $\chisqos$ only marginally.
Clearly, the octant discrimination becomes more and more difficult as
the true value of $\sa$ moves close to the maximal mixing. 
A combination of atmospheric as well as long-baseline experiments 
is needed to make this 
measurement \cite{Chatterjee:2013qus,Ghosh:2013pfa,Choubey:2013xqa,IoP-HRI-preparation}.


\section{Summary and Concluding Remarks}
\label{sec:conclusions}

The main aim of the upcoming ICAL experiment at INO is to determine
the mass hierarchy of neutrinos by observing the atmospheric neutrinos
and exploiting the Earth matter effects on their oscillations.
The major advantage of the ICAL detector is that it is well-tuned 
to detecting muons in the GeV range with a high efficiency,
and reconstructing their energy and direction with a high precision. 
It can also identify the charge of the muons, which allows it to 
distinguish between an incoming $\nu_\mu$ and $\bar{\nu}_\mu$, 
a capability that is beyond the reach of other large atmospheric 
neutrino experiments.
Because of these features the focus of the analyses for determining
the physics reach of ICAL has so far been on exploiting the 
high-precision information on muon momenta.

However, a large detector like ICAL with its calorimetric properties is also 
capable of measuring the hadron energy, which may be parameterized in 
terms of the observable $E'_{\rm had} \equiv E_\nu - E_\mu$ 
through a hadron hit calibration procedure. 
In this paper, we present the enhancement in the physics
reach of this experiment brought in by taking into account the combined
information in muon momentum and hadron energy in each event. We focus on
the identification of mass hierarchy and the precision measurements of 
atmospheric neutrino mixing parameters. 

The additional information we seek is contained not only in the hadron energy 
distribution of events, but also in the correlations between the hadron 
energy and muon momentum in each event. 
For example, by using both $E'_{\rm had}$ and $E_\mu$ as 
observables, we indirectly probe the incoming neutrino energy. 
However a naive addition of these two to reconstruct the neutrino 
energy would lose the advantage of precise muon energy determination,
and hence our analysis goes beyond that, by treating both these
observables separately for each event.  
Indeed, when the muon and hadron information 
is combined on an event-by-event basis, one also gets access to what 
fraction of energy of the incoming neutrino is carried by the muon.
This correlated muon and hadron information is what we try to
extract in this analysis.

We adopt a binning scheme in the observables 
($E_\mu$, $\cos\theta_\mu$, $E'_{\rm had}$),
where we divide the events in 10 $E_\mu$ bins, 21 $\cos\theta_\mu$ bins,
and 4 $E'_{\rm had}$ bins. We have used a relatively coarse
binning scheme since we would like to have sufficient 
number of events in all bins. 
Since the hadron energy resolution at ICAL is not as precise as 
that for the muon, the number of $E'_{\rm had}$ bins has been chosen 
to be small. The non-uniform bins are such that
the features relevant to the oscillations and matter effects are retained.
We demonstrate that such an analysis yields marked 
improvements over the analyses that use muon information alone.

Adding the hadron energy information to the muon information,
we find a significant enhancement in the capability of ICAL
to determine the neutrino mass hierarchy. For the benchmark
values of oscillation parameters, which are close to the current best-fit 
values, ICAL can determine the neutrino mass hierarchy with a significance 
of $\chisqmh \approx 9$ with 500 kt-yr exposure. 
This is an improvement of  more than $40\%$ over the analysis that uses 
only muon information. This also implies that the same value of 
$\chisqmh$ can be achieved with 40\% less exposure when the 
correlated hadron information is added.
We have also checked that the results with hadron energy are superior 
by about 35\% even when a finer binning scheme --- 20 $E_\nu$ bins
and 80 $\cos\theta_\mu$ bins --- is adopted for the analysis that uses
only muon information.
Depending on the true values of the oscillation parameters, the 
$\chisqmh$ value varies between 7 and 12, for an exposure of 500 kt-yr.
This is a crucial improvement, given that the main aim of the
ICAL experiment is the identification of mass hierarchy.

We also demonstrate that the atmospheric neutrino mixing parameters
can be measured more precisely by the inclusion of hadron energy information.
Addition of hadron energy information improves the $\sin^2 \theta_{23}$
precision marginally from 14\% to 12\%. 
However the precision on $|\mam|$ improves remarkably, from 4.2\%
to 2.9\%, even when the former has been obtained with the finer
binning mentioned above. 
This may be attributed to a better determination of the original neutrino
energy, and hence a better determination of the leading term in the
muon survival probability that oscillates as $\sin^2(\mam L/E_\nu)$. 
With the inclusion of hadron energy information, the expected precision 
on $|\mam|$ from ICAL after 500 kt-yr exposure is much better than the 
current reach of Super-Kamiokande; it is comparable to that obtained from 
MINOS, or with the current T2K data.

As far as the discrimination of $\theta_{23}$ octant is concerned, the hadron 
information increases the range of true $\theta_{23}$ values for which,
say, a $2\sigma$ discrimination is possible. However, such a discrimination
would be possible with 500 kt-yr exposure only with true NH and 
and $\sa (\text{true)}<0.395$. For higher values
of $\theta_{23}$, the reach of ICAL alone is still limited. 

It is clear from the above results that the inclusion of correlated 
hadron energy information improves oscillation physics sensitivities 
in almost all areas. However a few caveats are in order while
interpreting the final numbers. We use the ICAL detector response to 
the muons and hadrons, as obtained by the INO collaboration. 
Given the current status of the understanding of the ICAL detector response
obtained from simulations, we have had to make certain assumptions.
For example, we assume that the muon track and the hadron shower can be
separated completely in all events. We also neglect the background hits, 
noise, multiple hits, and assume that they do not affect the hadron
response of the detector. As the understanding of the detector improves, 
including the reconstruction for muons and hadrons and the separation of 
hits due to them, the physics reach could be affected. 
However, this paper demonstrates quantitatively that, with the
same conditions and assumptions, the inclusion of event-by-event hadron 
energy information in the analysis increases the reach for mass hierarchy 
identification and $|\mam|$ precision by a significant amount.

We expect this analysis procedure to become the preferred one for future 
analyses of ICAL physics reach. However it still does not extract all
possible information contained in the events, for example the
information in the hit pattern of hadron shower remains unexploited. 
A better understanding of the hadron response of the detector, 
and development of algorithms to use the hadron data efficiently
would be crucial in making the most of the data that would be available
from ICAL.

\subsubsection*{Acknowledgments}

This work is a part of the ongoing effort of INO-ICAL collaboration 
to study various physics potentials of the proposed ICAL detector. 
Many members of the collaboration have contributed for the completion 
of this work. We are very grateful to K. Bhattacharya, G. Majumder and 
A. Redij for their developmental work on the ICAL detector simulation package. 
In addition, we thank S. Choubey, A. Ghosh, N. Mondal, D. Indumathi, S. Uma Sankar, 
S. Goswami, N. Sinha, P. Ghoshal, and M. Naimuddin for 
intensive discussions on the oscillation analysis in regular meetings.
We would also like to thank V.M. Datar, R. Gandhi, and A. Raychaudhuri
for their useful comments on the paper. S.K.A would like to acknowledge M. Honda
for useful communications.
M.M.D. acknowledges the support from the Department of Atomic Energy (DAE) 
and the Department of Science and Technology (DST), Government of India.
T.T. thanks Costas Andreopoulos for discussions on the neutrino event 
generators, and Tata Institute of Fundamental Research for their extended 
support. S.K.A. acknowledges the support from DST/INSPIRE Research Grant 
[IFA-PH-12], Department of Science and Technology, India. 
A.D. acknowledges partial support from the European
Union FP7 ITN INVISIBLES (Marie Curie Actions, PITN-GA-2011-289442).

\bibliographystyle{JHEP}
\bibliography{ical-references}

\end{document}